\documentclass{aa}  

\usepackage{graphicx}
\usepackage{txfonts}
\usepackage{hyperref}
\usepackage{subfigure}
\usepackage{ragged2e}
\usepackage{xcolor}

\usepackage{microtype} 
\microtypesetup{protrusion=true,expansion=true}

\begin{document}

   \title{Magnetic fields at the dawn of structure formation}
   \subtitle{I. The CARLA J1510+5958 proto-cluster}
   \authorrunning{}

   \author{A. Pagliotta
          \inst{1}\fnmsep\inst{2}
          \and
          A. Bonafede\inst{1}\fnmsep\inst{2}
          \and
          C. Stuardi\inst{2}
          \and
          C. J. Riseley\inst{3}\fnmsep\inst{4}\fnmsep\inst{2}
          \and
          D. Vallés-Pérez\inst{1}\fnmsep\inst{2}
          \and 
          P. Tozzi\inst{5}
          \and
          L.\ Di Mascolo\inst{6}
          }
   \institute{Dipartimento di Fisica e Astronomia, Università di Bologna, via Piero Gobetti 93/2, I-40129 Bologna, Italy\\
              \email{annalisa.pagliotta@unibo.it}
         \and 
         INAF - Istituto di Radioastronomia di Bologna, Via Piero Gobetti 101, I-40129 Bologna, Italy
         \and
         Astronomisches Institut der Ruhr-Universit\"{a}t Bochum (AIRUB), Universit\"{a}tsstra{\ss}e 150, 44801 Bochum, Germany
        \and 
        Ruhr Astroparticle and Plasma Physics Center (RAPP Center), 44780 Bochum, Germany
        \and
        INAF - Osservatorio Astrofisico di Arcetri, Largeo E. Fermi 5, 50125 Firenze, Italy
        \and
        Kapteyn Astronomical Institute, University of Groningen, Landleven 12, 9747 AD, Groningen, The Netherlands
        }

   \date{Received Month Date, Year; accepted Month Date, Year}
 
  \abstract
   {Magnetic fields are a fundamental ingredient of the Universe; they influence the formation and evolution of cosmic structures. While magnetic fields in local galaxy clusters have been studied in a handful of cases, their origin, amplification, and strength at high redshift remain poorly understood. Proto-clusters represent the early stages of galaxy cluster formation, making them ideal laboratories to investigate the early magnetisation of the intracluster medium (ICM).}
   {We present a detailed study of CARLA J1510+5958 proto-cluster at $z=1.72$, observed with the JVLA in the L band (1–2 GHz). Our aim was to investigate the strength and structure of the magnetic field in the proto-ICM, and to understand the role of active galactic nuclei (AGN) in the magnetisation of the environment during early cluster formation.}
   {We analysed the Faraday rotation effect on the polarised emission from the central radio-loud AGN jets and lobes using the rotation measure (RM) synthesis and QU fitting technique. By mapping the RM and fractional polarisation, we inferred the magnetic field configuration along the line of sight. By modelling the QU spectra, we investigated the contributions from the AGN lobes and the surrounding medium. We further interpreted the observations with 3D simulations modelling gas density and turbulent magnetic fields, varying AGN orientation and path length through the magnetised plasma.}
   {The two lobes of the AGN show different polarisation properties. The western lobe exhibits a uniform RM (average $-115 \pm 32 \ \rm rad \ m^{-2}$) with low RM dispersion ($36 \pm 11 \ \rm rad \ m^{-2}$), indicating a partially ordered magnetic field, while the eastern lobe is depolarised. The polarisation asymmetry can be attributed to the presence of a turbulent, magnetised medium. However, simulations indicate that an isotropic, random turbulent magnetic field cannot reproduce the RM distribution observed in the western lobe, where the magnetic field is likely to be locally ordered, possibly due to compression induced by the lobe itself. The QU fitting further suggests an internal Faraday component, interpreted as magnetised relativistic plasma from the lobe mixed with the surrounding gas, indicating a possible magnetisation of the ambient medium by the AGN. Using the difference in polarisation between lobes, we constrain the average physical magnetic field strength in the proto-ICM to a lower limit of $0.4 \ \mu$G, in agreement with the only other estimate in a proto-cluster, which is an upper limit of 1 $\mu$G.} 
   {Our results confirm the presence of a magnetised plasma in the proto-ICM at $z = 1.72$, indicating early magnetic field amplification during cluster assembly. This study paves the way for further exploration of magnetism in the high-redshift Universe.}


   \maketitle

\section{Introduction} \label{Introduction}
In the local Universe, magnetic fields (MFs) are observed to permeate the intracluster medium (ICM) of galaxy clusters, as revealed by the presence of diffuse radio synchrotron emission extending up to megaparsec scales (e.g. \citealt{van_Weeren_2019} for a review). Despite their ubiquity, MFs have been studied thus far only in a limited number of clusters (e.g. \citealt{Bonafede_2010}, \citeyear{Bonafede_2013}; \citealt{Stuardi_2021}; \citealt{Osinga_2022}, \citeyear{Osinga_2025}; \citealt{DeRubeis_2024}; \citealt{Pagliotta_2025}). Current measurements indicate that the MF intensity reaches levels of a few microgauss throughout the cluster volume, with the highest values typically ranging from $1-12 \ \mu$G in the central regions. Observations also reveal that the MFs fluctuate spatially on scales of tens to hundreds of kiloparsecs. This turbulent nature is thought to arise from adiabatic compression and small-scale dynamo amplification of weak seed MFs, triggered by structure formation, accretion, and mergers (e.g. \citealt{Vazza_2014}, \citeyear{Vazza_2017}, \citeyear{Vazza_2018}; \citealt{Donnert_2018}; \citealt{DF_2019}; \citealt{Steinwandel_2022}). The origin of these fields remains an open question. They may originate in the early Universe (primordial scenario; e.g. \citealt{Subramanian_2016} for a review) or be injected by later-epoch astrophysical sources, such as galactic outflows or active galactic nuclei (AGN) (astrophysical scenario; e.g. \citealt{Donnert_2009}; \citealt{Xu_2011}). Because the small-scale turbulent dynamo in galaxy clusters reaches saturation (i.e. equipartition between the MF energy and the flow kinetic energy), it yields a similar level of MF amplification across different systems. Therefore, both primordial and astrophysical scenarios can, in principle, explain the observed MF strength. However, recent works disfavour a purely astrophysical origin (e.g. \citealt{Carretti_2022}, \citeyear{Carretti_2023}, \citeyear{Carretti_2025}). Despite these insights, no clear correlation has yet been established between the MF properties and the dynamical state of the clusters, likely due to limited statistics. Nonetheless, hints of correlations have been reported between the total MF energy and cluster mass \citep{DF_2019}, and between central MF intensity (i.e. the average MF in the cluster core) and central gas density \citep{Govoni_2017}. 

Observations of galaxy clusters at intermediate redshifts (e.g. \citealt{Di_Gennaro_2020}) have contributed to our understanding of the MF amplification, showing that microgauss-level MFs can already be in place at $z \sim 0.6-0.9$. This result is unexpected, as it implies a faster amplification than commonly assumed. If MFs are amplified primarily by small-scale dynamos driven by merger-induced turbulence, saturation is expected a few gigayears after cluster formation (e.g. \citealt{Vazza_2018}). The observed field strengths may therefore suggest additional amplification mechanisms that contribute alongside dynamos. We note, however, that this analysis relies solely on total intensity measurements and cannot disentangle the relative contributions of cosmic rays and MFs to the observed radio brightness. Recent observations at high redshift ($z>1$) support this fast MF amplification scenario through the detection of diffuse radio emission in galaxy clusters (e.g. \citealt{Phuravhathu_2025}; \citealt{DiGennaro_2025}), including the most distant radio halo at $z=1.23$ \citep{Sikhosana_2025}, and a candidate mini-halo at $z=1.709$ \citep{HLavacek_Larrondo_2025}. Hence, to fully understand the origin and evolution of MFs in cluster environments, it is essential to probe even earlier phases, namely proto-clusters, where MFs and their amplification processes remain largely unexplored.\\    
\\
Proto-clusters are considered the high-redshift progenitors of local galaxy clusters (see \citealt{Overzier_2016} for a review), and are generally identified as galaxy overdensities associated with dark matter (DM) haloes that will eventually merge and collapse into massive ($\rm M \gtrsim 10^{14} \ M_\odot$) clusters \citep{Muldrew_2015}. However, the term proto-cluster encompasses a wide range of evolutionary stages \citep{Muldrew_2015}, where those identified at $z \sim 7 - 8$ (e.g. \citealt{Hu_2021}; \citealt{Morishita_2023}) are completely distinct complexes than the ones observed at later epochs. At $z=2$, proto-clusters can span tens of comoving megaparsecs (e.g. \citealt{Chiang_2013}, \citeyear{Chiang_2015}, \citeyear{Chiang_2017}), often hosting a large fraction of dusty star‑forming galaxies (e.g. \citealt{Cheng_2019}; \citealt{Zhang_2024}), and enhanced AGN activity (e.g. {\citealt{Tozzi_2022a}; \citealt{Vito_2024}; \citealt{Travascio_2025}). In addition, proto-clusters and high-redshift clusters represent distinct phases of structure formation, as high‑redshift clusters ($z>1$) are more compact, are dominated by a single massive halo, and exhibit a reduced AGN fraction and a more well-established red sequence (e.g. \citealt{Strazzullo_2019}; \citealt{Willis_2020}; \citealt{Leste_2024}).

Simulations predict that significant thermal energy is already present in proto-clusters as early as $z \sim 2-4$ \citep[e.g.][]{Barnes_2017, Remus_2023}, implying that ICM assembly begins in these early structures. However, detecting the proto-ICM through X-ray or Sunyaev--Zeldovich (SZ) observations remains challenging. There is starting evidence for the presence of large-scale reservoirs of hot gas (e.g. \citealt{Dong_2023}), as well as detections in a handful of more compact cores (e.g. \citealt{Zhou_2026}). Recent works have also reported thermal gas in the Spiderweb proto-cluster at $z=2.16$ through both X-ray and SZ observations (\citealt{Tozzi_2022b}; \citealt{Di_Mascolo_2023}; \citealt{Lepore_2024}). To date, this object represents the only proto-cluster environment where the proto-ICM MF has been directly investigated; \citet{Anderson_2022}, using Jansky Very Large Array (JVLA) polarisation data in the X ($8-12$ GHz) and Ka ($29.2-36.8$ GHz) bands, estimated an upper limit of $1 \ \mu$G for the physical MF intensity and suggested that the central radio-loud AGN actively contributes to the magnetisation of the surrounding medium. 
\\
\\
The most effective strategy for probing a magnetised medium (e.g. the ICM) is to observe radio galaxies, which are strong sources of linearly polarised radiation from their jets and lobes, either embedded within or located behind the medium of interest. Linear polarisation is described by the Stokes parameters $Q$ and $U$, combined as
\begin{equation}
    \rm \mathbf{P} = Q + iU = pI e^{2i\psi}, \ \ \ \
    \psi = \frac{1}{2} \arctan \Biggl(\frac{U}{Q}\Biggr),
    \label{eq:P}
\end{equation}
where $p$ is the fractional polarisation and $I$ is the total intensity. As the radiation propagates through a magnetised, ionised medium, the polarisation angle $\psi$ is rotated (Faraday rotation; \citealt{Burn_1966}), such that $\psi_{\rm obs} = \psi_0 + {\rm RM}\cdot \lambda^2$, where RM is the rotation measure, defined as
\begin{equation}
    \rm {\rm RM} = 812 \int_{z}^{0} \frac{n_e(z)\ B_\parallel(z)}{(1+z)^2} \frac{dl}{dz}\ dz 
    \ {\rm rad\ m^{-2}}.
    \label{eq:RM}
\end{equation}
Here $n_e$ is the physical electron number density in $\rm cm^{-3}$, $B_\parallel$ is the component of the physical MF along the line of sight (LOS) in $\mu$G,\footnote{The comoving MF is obtained by dividing the physical MF by $(1+z)^2$.} $dl$ is the proper path length in kiloparsecs, and $z$ is the redshift of the Faraday rotating medium. The $(1+z)^2$ factor accounts for redshift dependence of the observed wavelength.

Various state-of-the-art techniques have been developed to derive the RM from radio polarised observations, such as the RM synthesis technique (\citealt{Brentjens_2005}) and the QU fitting technique (\citealt{OSullivan_2012}), and have been extensively applied in galaxy clusters studies (e.g. \citealt{Stuardi_2021}; \citealt{DeRubeis_2024}; \citealt{Pagliotta_2025}; \citealt{Di_Gennaro_2021}, \citeyear{Di_Gennaro_2023}; \citealt{Rajpurohit_2022}). Then, to determine the MF properties, such as the intensity, profile, and power spectrum (i.e. how the MF energy is distributed across different scales), one approach consists in reproducing the ICM or proto-ICM through semi-analytical 3D simulations and creating mock RM maps to compare with the observations. These simulations account for the gas density profiles derived from X-ray or SZ observations (if available), and assume a MF power spectrum. Early studies used a Kolmogorov power-law model (e.g. \citealt{Murgia_2004}; \citealt{Bonafede_2010}; \citealt{Vacca_2010}), whereas recent works (\citealt{Stuardi_2021}; \citealt{DeRubeis_2024}; \citealt{Pagliotta_2025}) have adopted more realistic models from magnetohydrodynamic (MHD) simulations of galaxy clusters that include the small-scale dynamo amplification (\citealt{DF_2019}).\\
\\
Proto-clusters hosting central radio-loud AGN (e.g. \citealt{Hatch_2014}; \citealt{Anderson_2022}; \citealt{Chapman_2024}) are particularly well suited for this type of study as the polarised emission from the jets and lobes of the host radio galaxy provides a perfect embedded source to probe the surrounding magnetised medium through Faraday rotation. One of the largest samples of such systems is provided by the Clusters Around Radio-Loud AGN (CARLA) survey (\citealt{Wylezalek_2013}, \citeyear{Wylezalek_2014}). This survey targeted 387 powerful radio-loud AGN (187 quasars and 200 high-redshift radio galaxies) in the redshift range $1.3 < z < 3.2$, using mid-infrared imaging of the Spitzer space telescope. Approximately 55\% (10\%) of the CARLA fields show significant galaxy overdensities above the 2$\sigma$ (5$\sigma$) level, relative to the Spitzer UKIDSS Ultra Deep Survey. The spectroscopically confirmed overdense regions have halo masses of $13.5 \lesssim \rm \log_{10}(M_{halo}/M_\odot) \lesssim 14.5$ (\citealt{Mei_2023}; \citealt{Afanasiev_2023}), consistent with progenitors of present-day clusters, and often exhibit multiple components, as predicted for early cluster formation (\citealt{Chiang_2013}; \citealt{Muldrew_2017}). Follow-up observations with the Hubble space telescope (\citealt{Noirot_2016}, \citeyear{Noirot_2018}) confirmed 16 of these overdensities as likely clusters at $1.3 \lesssim z \lesssim 2.8$, along with seven additional serendipitous structures at $0.9 < z < 2.1$. Recently, \citealt{Grishin_2024} spectroscopically confirmed an additional CARLA cluster at $z=2.36$ and serendipitously discovered a new one at $z=2.24$. These structures have been widely studied in optical and infrared wavelengths to investigate galaxy formation and evolution, star formation activity, and the growth of supermassive black holes (e.g. \citealt{Marinello_2020}; \citealt{Mei_2023}; \citealt{Afanasiev_2023}).\\
\\
Using the JVLA, we have started the first systematic study of the MF in proto-clusters selected from the CARLA survey. Out of 25 of the spectroscopically confirmed CARLA structures mentioned above, our final sample comprises nine targets. We excluded proto-clusters hosting AGN with weak polarisation ($< 2 \sigma_{\rm RMS}$) in the NRAO VLA Sky Survey (NVSS; \citealt{Condon_1998}), and those not well resolved in total intensity at $3^{\prime\prime}-5^{\prime\prime}$ based on both Faint Images of the Radio Sky at Twenty-cm (FIRST;\footnote{Available at \url{https://sundog.stsci.edu/index.html}} \citealt{Becker_1995}) or VLA Sky Survey (VLASS;\footnote{Available at \url{https://science.nrao.edu/vlass}} \citealt{VLASS_2020}) images. Furthermore, we excluded fields with high Galactic RM uncertainty ($ \rm >10 \ rad \ m^{-2}$) based on the \citealt{Hutschenreuter_2022} Galactic Faraday rotation map. In addition, three targets were rejected during the telescope's technical review, and two new discoveries were confirmed only after our observations had already been scheduled.

In this paper we present the initial results for one of these systems, CARLA J1510+5958 ($\rm RA_{J2000} = 15h \ 10m \ 05s$, $\rm DEC_{J2000}= +59^{\circ} \ 58' \ 55''$; \citealt{Condon_1998}), characterised by a halo mass of $\rm log_{10}(M_{halo}/M_{\odot}) = 14.2$ \citep{Mei_2023} and centered around a radio-loud AGN at $z=1.719$ \citep{Wylezalek_2013}. To date, six members of this system have been spectroscopically confirmed \citep{Noirot_2018}. The analysis of the remaining eight proto-clusters will be presented in a forthcoming study (Pagliotta et al., in prep.). The aim of this work is to constrain the properties of the MF in these early structures, which trace the very first phases of structure formation and remain nearly unexplored in this context. By exploiting JVLA L band ($1-2$ GHz) observations of the central radio-loud AGN, our aim was to determine the MF strength and configuration in the proto-ICM. These observations also allowed us to investigate the eventual contribution of the central source to the magnetisation of the surrounding environment. We processed and imaged the data in both total intensity and polarisation. The data reduction process is summarised in Sect. \ref{Data reduction and analysis}, along with an overview of the RM synthesis and the QU fitting technique. In Sect. \ref{RMS results} we present the results of our analysis, including measurements of the RM, its dispersion ($\sigma_{\rm RM}$), and the fractional polarisation, while in Sect. \ref{QU fitting results} we provide the results of the QU fitting. Section \ref{MF estimate} describes the methods used to constrain the properties of the MF, including the comparison of simulations with the observations. A detailed discussion of these results is provided in Sect. \ref{Discussion}, followed by a summary of the study in Sect. \ref{Conclusions}. Additional details of our calibration and analysis are provided in Appendices \ref{3C286 model}, \ref{QU fitting-2}, and \ref{MockRM}.

Throughout the paper a Lambda cold dark matter ($\Lambda$CDM) cosmology is assumed, with $\rm H_0 = 70 \ km \ s^{-1} \ Mpc^{-1}$, $\rm \Omega_M = 0.3$, and $\Omega_{\Lambda} = 0.7$. This translates to a luminosity distance of $\rm D_L$ = 12911.9 Mpc and a scale of 8.461 kpc/arcsec at the proto-cluster redshift, z = 1.719.

\section{Data reduction and analysis} \label{Data reduction and analysis}
\subsection{Data specifics and calibration} \label{Calibration}
CARLA J1510+5958 (hereafter CARLAJ1510) was observed with the JVLA under the project 24A-040 (PI: Bonafede). Observations were carried out in full polarisation in the L band frequency range ($1-2$ GHz), using the B-array configuration, which provides a resolution of $\sim 4^{\prime\prime}$ at 1.5 GHz.\footnote{See \url{https://science.nrao.edu/facilities/vla/docs/manuals/oss2026a/performance/resolution}} The total on-source integration time was 3h 16m, resulting in a theoretical sensitivity of $\rm \sim 6 \ \mu Jy \ beam^{-1}$ in both total intensity and polarisation. This source was selected for initial analysis as it was observed during nighttime, reducing radio frequency interference (RFI) and mitigating contamination from elevated solar activity, which was at its maximum during the observation period (July 2024).

\begin{figure*}
    \centering
    \hspace{-0.42cm}
    \begin{subfigure}
        \centering
        \includegraphics[width=0.339\linewidth]{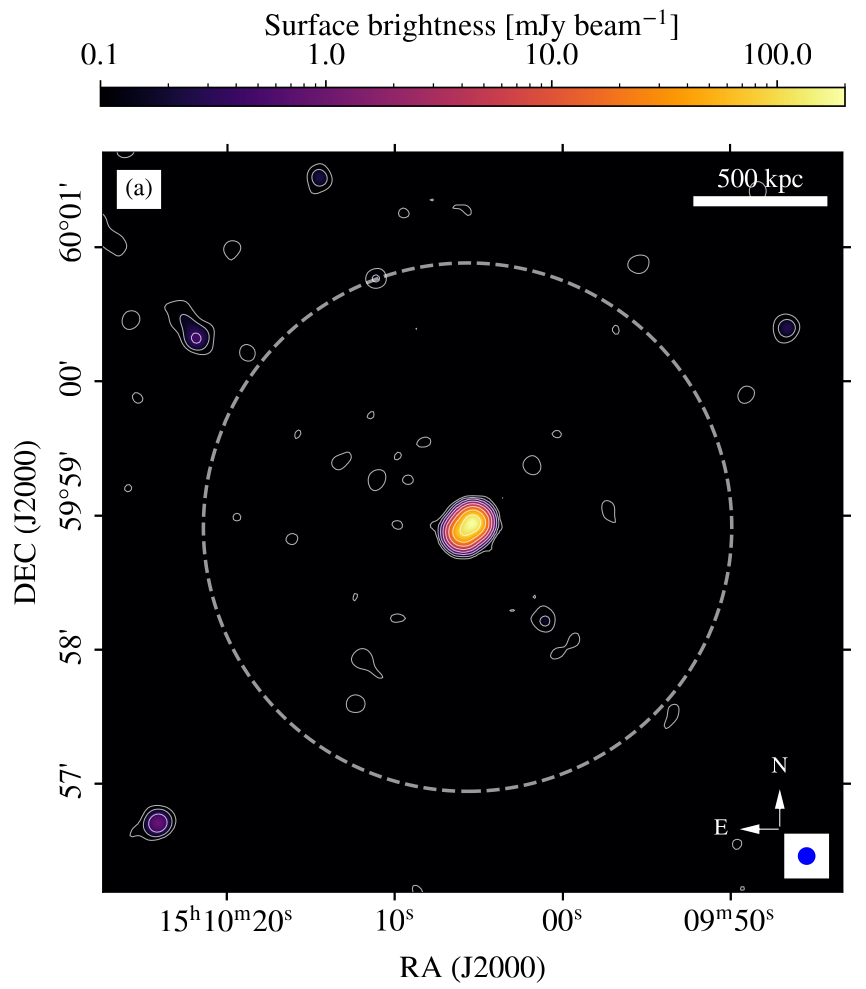}
    \end{subfigure}
    \hspace{-0.3cm}
    \begin{subfigure}
        \centering
        \includegraphics[width=0.342\linewidth]{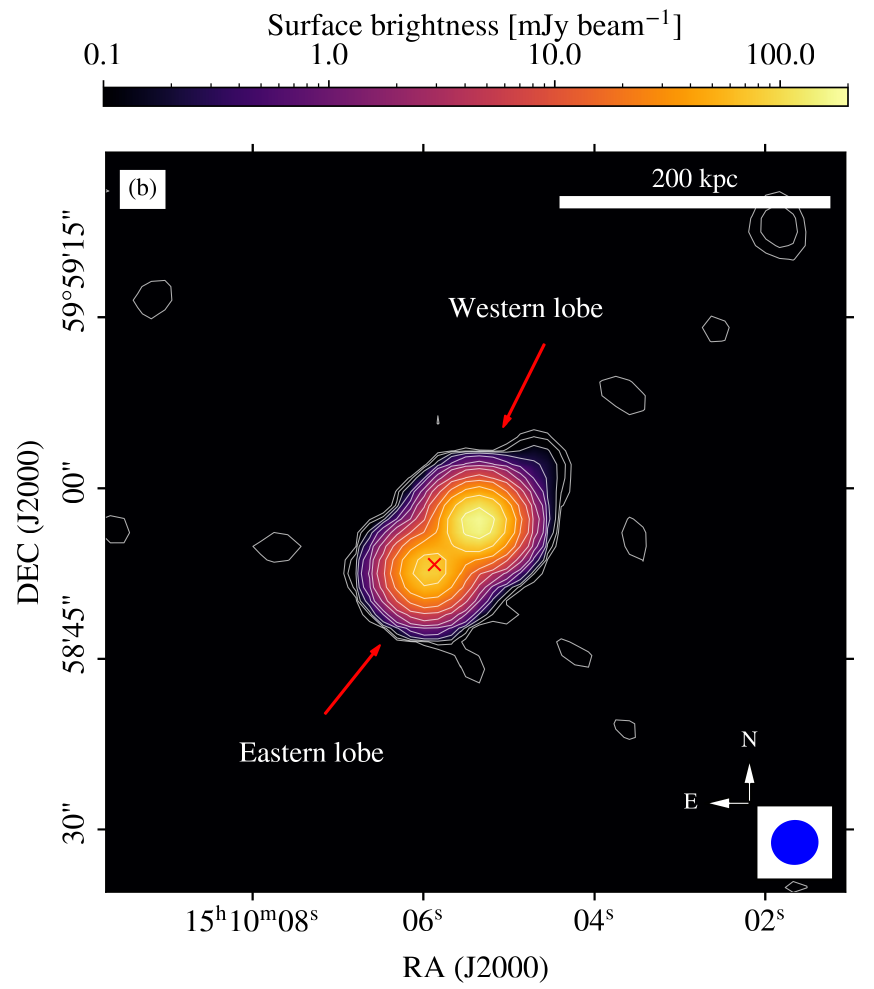}
    \end{subfigure}
    \hspace{-0.42cm}
    \begin{subfigure}
        \centering
        \includegraphics[width=0.3485\linewidth]{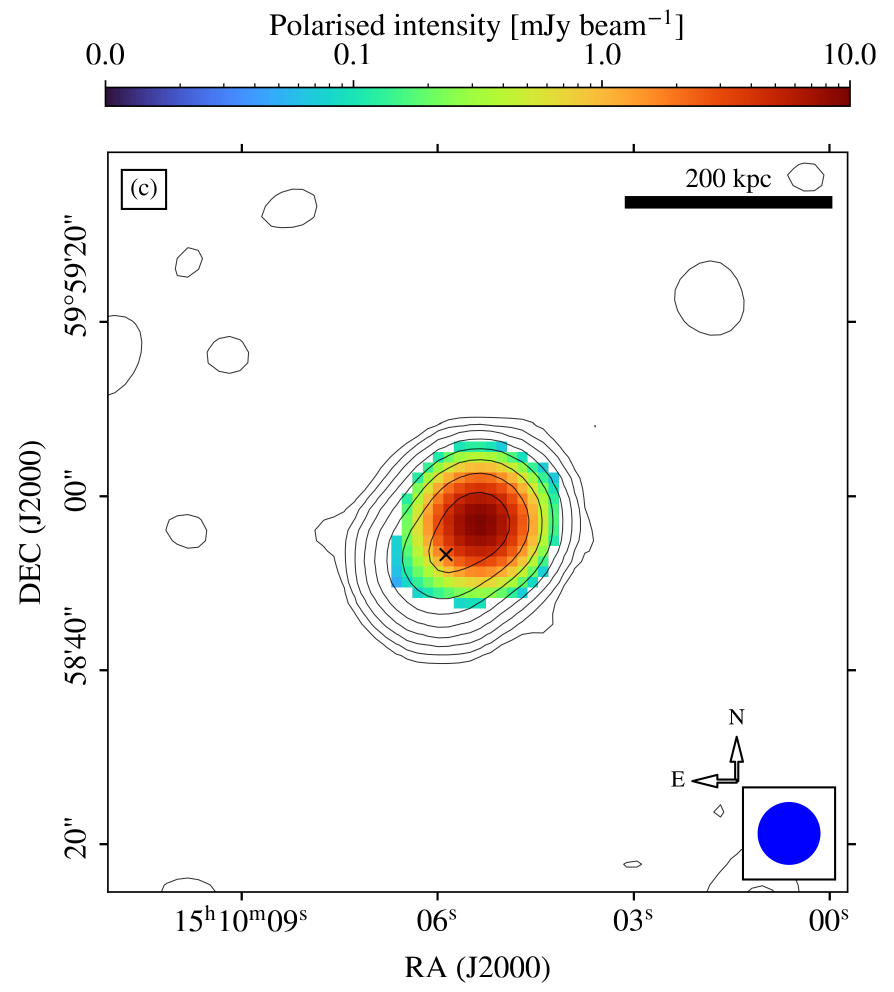}
    \end{subfigure}
    \vspace{-0.2cm}
    \caption{Total intensity and polarisation maps of CARLAJ1510 and central source. Panel (a): Stokes I map of CARLAJ1510 and surrounding field at 1.5 GHz, produced using Briggs weighting (robust = 0.5). The dashed circle corresponds to the reference distance of 1 Mpc from the proto-cluster centre. Panel (b): Zoomed-in total intensity map of the central radio-loud AGN, shown at the original resolution before convolution. The western and eastern lobes are indicated with the red arrows. Panel (c): Map of the maximum polarised intensity of the central source. Values were corrected for the Ricean bias. Only pixels above the $6\sigma_{\rm P,i}$ threshold are shown. Radio contours at 1.5 GHz (robust = 0.5) are overlaid in panels (a) and (c), starting from $3\sigma_{\rm I}$ (with $\sigma_{\rm I}=14.3 \ \mu \rm Jy \ beam^{-1}$) and scaling by a factor of 3. In panel (b), the contours start from  $3\sigma_{\rm I}$ (with $\sigma_{\rm I}=10.0 \ \mu \rm Jy \ beam^{-1}$) and scale by a factor of 2. The optical position of the host galaxy is indicated by the cross (\citealt{Gaia_2021}). The spatial scales and resolution beams are reported on the edges of the images: $7^{\prime\prime} \times 7^{\prime\prime}$ ($\sim 59$ kpc) for panels (a) and (c), and $\sim 4^{\prime\prime} \times 4^{\prime\prime}$ ($\sim 34$ kpc) for panel (b).}
    \label{fig:StokesI+P}
    \vspace{-0.2cm}
\end{figure*}

We first processed the data in continuum, followed by polarisation calibration using the Common Astronomy Software Application (\texttt{CASA}). The calibration sources used in the reduction step were 3C286 for parallel and cross-hand delay calibration, bandpass calibration, absolute flux density scale (\citealt{Taylor_2024}), and polarisation angle calibration; J1438+6211 for time-dependent complex gain calibration; and J1407+2827 for on-axis polarisation leakage calibration. Recently, \citet{Taylor_2024} and \citet{Hugo_2024}\footnote{\url{https://doi.org/10.48479/bqk7-aw53}} presented updated frequency-dependent physical models of the radio source 3C286 based on MeerKAT observations. In particular, \citet{Hugo_2024} proposed a new low-frequency model describing the fractional polarisation and polarisation angle (see Sect. 4.4 of \citet{Hugo_2024} for details). To summarise, within the L band frequency range, the intrinsic polarisation angle of 3C286 (after ionospheric correction) deviates significantly from the commonly assumed 33° value extrapolated from higher frequencies (above C and X bands), with RM varying slightly from 0 to 0.3 $\rm rad \ m^{-2}$ at the bottom of the band. Therefore, we proceeded as follows:
\begin{itemize}
    \itemsep2pt
    \item We ran the VLA continuum calibration pipeline,\footnote{Available at \url{https://science.nrao.edu/facilities/vla/data-processing/pipeline}} keeping the automatic RFI flagging applied in order to start the manual calibration on flagged data.
    \item We replaced the default model of 3C286 in the L band from \citet{Perley_2017} with the updated models provided by \citet{Taylor_2024} for the total intensity, and by \citet{Hugo_2024} for the fractional polarisation and polarisation angle (see Appendix \ref{3C286 model} for details).
    \item We then performed two iterations of standard continuum calibration (parallel delay, bandpass, and complex gain calibration) and applied the corrections to the three calibrators only. Manual and automatic flagging were applied to both parallel-hand and cross-hand data.
    \item We used the updated model of 3C286 to solve for the cross-hand delay by combining the spectral windows. The on-axis polarisation leakage (D-terms) was then derived every 1 MHz using J1407+2827. Finally, we used 3C286 again to solve for the R–L polarisation angle, also at 1 MHz intervals.
    \item All the above polarisation calibration solutions were applied to the three calibrators, followed by flagging. The polarisation calibration steps were repeated twice to allow for flagging on the calibrated cross-correlations.
    \item Finally, we applied  the continuum and the polarisation calibration solutions to the target, with parallactic angle correction enabled, and performed additional flagging of the data.
\end{itemize}
Before applying the solutions to CARLAJ1510, we checked the new calibration by imaging 3C286 in all Stokes parameters (I, Q, U, and V) using \texttt{WSClean v3.6.0}\footnote{Available at \url{https://gitlab.com/aroffringa/wsclean}} (\citealt{Offringa_2014}, \citeyear{Offringa_2017}). We assessed whether the recovered Stokes I, fractional polarisation, and polarisation angle matched the input models.

To ensure that time-averaging smearing remains below a 1\% amplitude loss in the B-array configuration, we averaged the data in time using a bin size of six seconds, without averaging in frequency to avoid bandwidth smearing outside the primary beam. Self-calibration was performed using \texttt{CASA}, and imaging and model prediction were carried out with \texttt{WSClean v3.6.0}.
We applied four rounds of phase-only self-calibration, followed by three rounds of amplitude self-calibration. Subsequently, we performed a direction-dependent amplitude self-calibration on three off-axis sources and subtracted all the sources located outside the primary beam ($30^{\prime}$ at 1.5 GHz). As a final step, we performed one round of phase-only self-calibration to refine the gain corrections.
\vspace{-0.6cm}

\subsection{Imaging} \label{Imaging}
In preparation for the further analysis, images of Stokes Q and U were produced using \texttt{WSClean v3.6.0}. For the RM synthesis technique (see Sect. \ref{RMS}), the full bandwidth was divided into 112 channels of 8 MHz each, to reach the desired Faraday depth ($\phi$) sensitivity. Stokes I imaging included a second-order spectral fit and primary beam correction using the JVLA antenna model.\footnote{\url{https://everybeam.readthedocs.io/en/latest}} Stokes Q and U were imaged jointly with the options \texttt{-pol qu}, \texttt{-join-polarizations}, and \texttt{-squared-channel-joining}. A mask was generated from a multifrequency synthesis (MFS) Stokes I image built from 16 channels.
Imaging was carried out with a pixel size of 0.6$^{\prime\prime}$ over 3000 pixels, using the Briggs weighting scheme (\citealt{Briggs_1995}) with robust parameter of 0.5 to recover both compact and extended AGN emission. The Stokes Q and U images were then smoothed to a common  $7^{\prime\prime}\times7^{\prime\prime}$ ($\sim59$ kpc) circular beam to minimise beam depolarisation, using the \texttt{imsmooth} \texttt{CASA} task. Channels with a noise larger than 6 times the RMS noise over the median noise, as well as channels totally flagged or with a beam larger than 1.5 times the median beam, were excluded resulting in 98 usable channels. Finally, the cubes were re-binned for computational load by a factor of two in a final pixel scale of 1.2$^{\prime\prime}$ ($\sim 10$ kpc). We show in Fig. \ref{fig:StokesI+P}, panel (a) the Stokes I MFS image (robust = 0.5) after the convolution and the re-binning with RMS noise $\sigma_{\rm I}=14.3 \ \mu \rm Jy \ beam^{-1}$. Fig. \ref{fig:StokesI+P}, panel (b) shows a zoom-in of the central radio source at the original resolution of $\sim 4^{\prime\prime} \times 4^{\prime\prime}$ ($\sim 34$ kpc) before convolution. To focus the analysis on the proto-cluster region, we cropped the re-binned Stokes Q and U cubes to include only the central area within a $2^{\prime}$ radius around the main source. This scale, corresponding to $\sim 1$ Mpc at the source redshift (indicated by the circle in Fig. \ref{fig:StokesI+P}, panel (a)), was chosen based on \citet{Noirot_2018}, where six confirmed members lie within this region.

For the QU fitting analysis (see Sect. \ref{QU fitting}), the same imaging parameters were adopted, but the data were averaged in frequency over 32 MHz, resulting in 28 channels, in order to improve the signal-to-noise (S/N) ratio and reduce computational cost. The subsequent imaging processing followed the same procedure described above, producing frequency cubes for all Stokes parameters, with all images convolved to a common resolution of $8\arcsec \times 8\arcsec$ ($\sim68$ kpc).
\vspace{-0.2cm}
\begin{figure*}[!htb]
    \vspace{-0.2cm}
    \centering
    \begin{subfigure}
        \centering
        \includegraphics[width=0.41\linewidth]{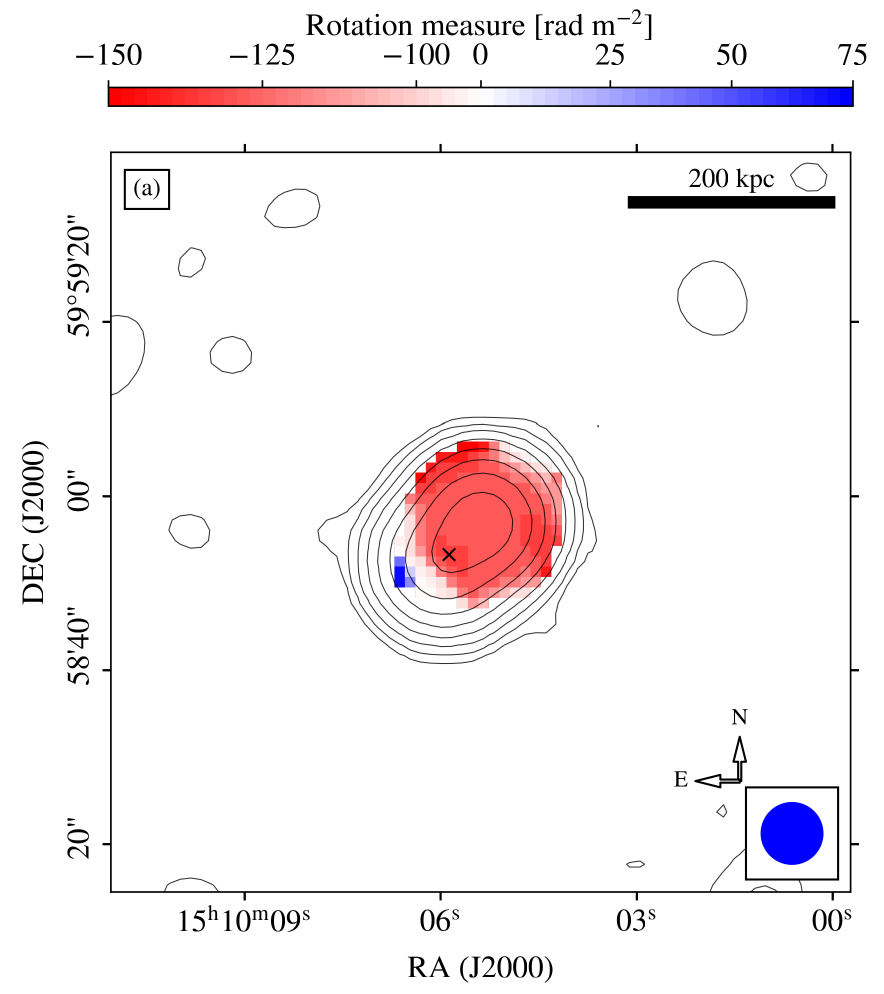}
    \end{subfigure}
    \hspace{1cm}
    \begin{subfigure}
        \centering
        \includegraphics[width=0.4075\linewidth]{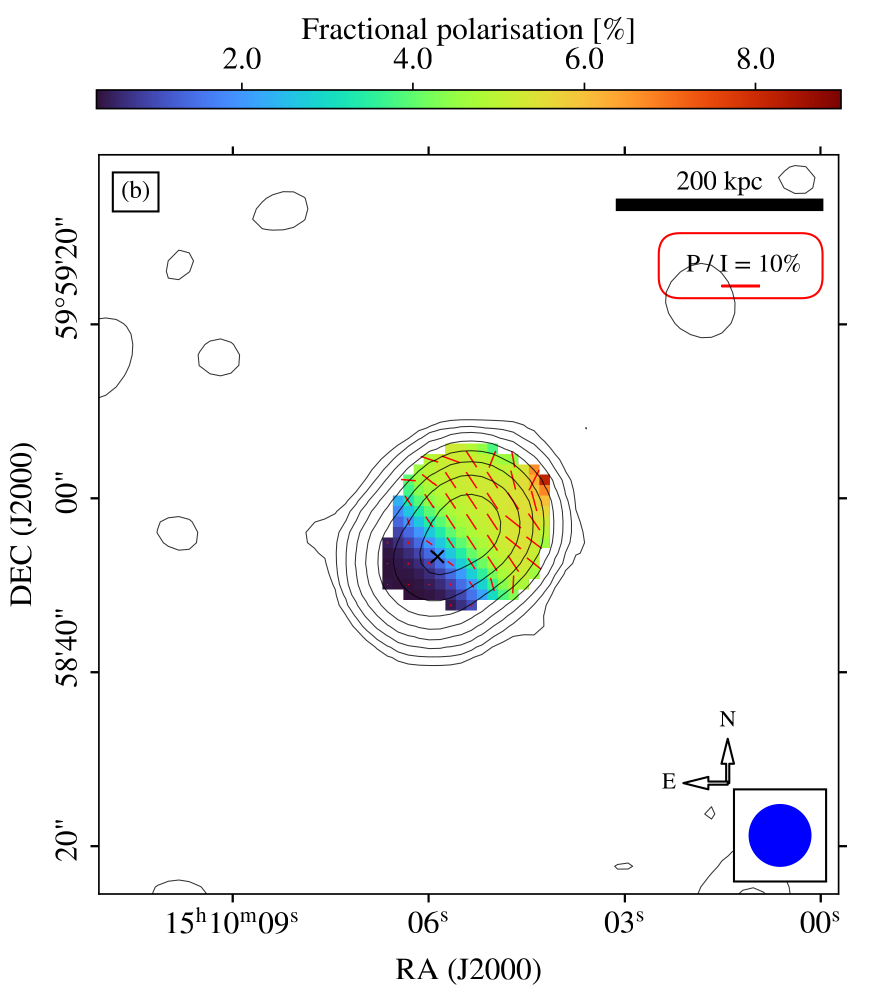}
    \end{subfigure}
    \vspace{-0.2cm}
    \caption{Rotation measure and fractional polarisation maps of the central source. Panel (a): RM map. Values were corrected for the Galactic foreground rotation and are in the source rest frame. Positive values refer to the MF orientation along the LOS towards the observer, whereas negative values refer to an orientation away from the observer. Panel (b): $\rm F_p$ map. Values were corrected for the Ricean bias. Magnetic field vectors at the source location are overlaid in red, with width scaling with $\rm F_p$. Only pixels above the $6\sigma_{\rm P,i}$ threshold are shown. Radio contours at 1.5 GHz (robust = 0.5), starting from $3\sigma_{\rm I}$ and scaling by a factor of 3 are overlaid in both images. The optical position of the host galaxy is indicated by the cross (\citealt{Gaia_2021}). The spatial scales and resolution beam of $7^{\prime\prime} \times 7^{\prime\prime}$ ($\sim 59$ kpc) are reported on the edges of the images.}
    \label{fig:RM+fracP}
    \vspace{-0.2cm}
\end{figure*} 
\subsection{RM synthesis technique} \label{RMS}
The main approach we adopted to estimate the RM is the RM synthesis technique developed by \citet{Brentjens_2005}. This method transforms the frequency-dependent polarised signal into Faraday space, allowing the reconstruction of the Faraday dispersion function (FDF) and the identification of multiple Faraday depth components along a single LOS. In the following, we refer to the Faraday depth, $\phi$, as the coordinate in Faraday space where the RM synthesis is carried out. The Faraday depth coincides with the RM at all wavelengths only when non-emitting media are present between the source and the observer.

Before the analysis, we computed key parameters that characterise the RM synthesis, which depend on the total bandwidth and the spectral resolution of the dataset (see \citealt{Brentjens_2005}). For our L band configuration, we derived:
\begin{itemize}
    \itemsep2pt
    \item[-] A maximum observable Faraday depth of $|\phi_{\mathrm{max}}| = 1322 \ \rm rad \ m^{-2}$, which corresponds to $9781 \ \rm rad \ m^{-2}$ in the rest frame of the proto-cluster.
    \item[-] A resolution in Faraday space, corresponding to the full width at half maximum (FWHM) of the rotation measure transfer function (RMTF), of $\delta\phi = 55 \ \rm rad \ m^{-2}$.
    \item[-] A maximum observable Faraday scale (i.e. the scale to which the instrument is sensitive at the 50\% level) of $\Delta\phi_{\mathrm{max}} = 143 \ \rm rad \ m^{-2}$.
\end{itemize}   

We applied the RM synthesis using the \texttt{rmsynth3d} task included in the Canadian Initiative for Radio Astronomy Data Analysis (\textsc{CIRADA}) \textsc{RM-Tools}.\footnote{Available at \url{https://github.com/CIRADA-Tools/RM-Tools}} The reconstructed FDF cubes span a Faraday depth range of $\pm \ 2000 \ \rm rad \ m^{-2}$ in steps of $1 \ \rm rad \ m^{-2}$, ensuring full coverage of the Faraday depth space and adequate sampling relative to the expected resolution. We subsequently used the \texttt{rmclean3d} task to perform a deconvolution of the dirty Faraday spectra, mitigating the effects of RMTF sidelobes. The cleaning threshold was set to $6\sigma$ ($851.2 \ \mu\rm Jy \ beam^{-1}$), which corresponds to a Gaussian significance level of $4.8\sigma$ according to \citet{Hales_2012}. This is based on the RMS noise estimated from the brightest polarised pixel and was calculated using a slice of the FDF spectrum in Faraday space, selected outside the sensitivity range ($|\phi| \geq 1500 \ \rm rad \ m^{-2}$), where no real signal is detected or expected. 
\vspace{-0.4cm}
\subsection{RM and polarisation maps} \label{Final maps}
\vspace{-0.2cm}
To produce the RM and polarisation maps, we first generated a noise map from the deconvolved total FDF cube, extending the previous method to each pixel, i, of the image, and used this to obtain a pixel-wise RMS estimate of the noise under a Gaussian approximation of the Ricean statistics. A $6\sigma_{\rm P,i}$ threshold (with $\langle \sigma_{\rm P,i} \rangle = 11.0 \rm \ \mu Jy \ beam^{-1}$) was then applied to create a mask, retaining only significant polarised detections. The polarisation map (Fig. \ref{fig:StokesI+P}, panel (c)) was then constructed from the CLEANed FDF cube by taking the peak polarised intensity in each pixel, $\rm |\tilde{F}(\phi_{\rm peak})|$, and correcting for Ricean bias by subtracting $2.3\sigma_{\rm P,i}^2$ in quadrature (\citealt{George_2012}). The RM map (Fig. \ref{fig:RM+fracP}, panel (a)) was derived from the peak Faraday depths ($\phi_{\rm peak}$),\footnote{Equivalent of the RM map in the case of Faraday-simple screens (i.e. characterised by a single value of $\phi$).} and corrected for Galactic foreground rotation, for which we adopted the median value of $3.2 \pm 3.7 \ \mathrm{rad} \ \mathrm{m}^{-2}$ (computed over a region of $0.4 \times 0.4 \ \rm deg^2$ from the \citealt{Hutschenreuter_2022} Galactic Faraday rotation map), and redshift effects using a $(1+z)^2$ factor (see Eq. \ref{eq:RM}), with $z = 1.72$. Finally, the fractional polarisation ($\rm F_p$) map (Fig. \ref{fig:RM+fracP}, panel (b)) was obtained by dividing the bias-corrected polarisation map by the Stokes I MFS image (Fig. \ref{fig:StokesI+P}, panel (a)) trimmed to match the size of the polarisation map.

\subsection{QU fitting} \label{QU fitting}
A complementary approach to study RM is the QU fitting (e.g. \citealt{OSullivan_2012}). Unlike the RM synthesis, which recovers the FDF via a Fourier transform, this method assumes a physical model for the magneto-ionic medium and fits the predicted polarisation spectrum to the observed Stokes Q and U spectra of the source as a function of $\lambda^2$ (Eq. \ref{eq:P}). This allows the identification of multiple Faraday components along the LOS, including both Faraday-simple and Faraday-complex structures, and whether the Faraday rotation occurs internally or externally to the source. By fitting to the data at each frequency channel, rather than reconstructing a Faraday depth spectrum, the QU fitting reduces depolarisation and blending of components close in Faraday depth. However, this analysis relies on assumptions about the underlying model.
This technique has been successfully applied to AGN and radio galaxies \citep[e.g.][]{OSullivan_2012, Anderson_2016, Kaczmarek_2018, Pasetto_2021}.

The QU fitting was performed with the \texttt{qufit} task from \textsc{CIRADA RM-Tools} \citep{Purcell_2020}, which applies Bayesian model fitting and supports a range of models (see Appendix \ref{QU fitting model}). The tool uses Stokes I, Q, and U flux densities (in Jy) as a function of frequency, with uncertainties, as input. The fitting is performed independently for each spatial element (i.e. source region) and no intrinsic spatial morphology is assumed, as the spatial response is set by the restoring beam. For each Stokes parameter $i$, the flux density $\rm F_i$ was extracted by selecting pixels above $\rm 3\sigma_I$. By default, the Stokes I spectrum was fitted using a second‑order logarithmic function and used to normalise the Stokes Q and U spectra. The output includes best fit parameter values with uncertainties, posterior distributions for all model parameters, and comparisons between the observed and modelled spectra.

\section{RM synthesis results} \label{RMS results}
Here, we present the results from the RM synthesis. The most relevant values are summarised in Table \ref{tab:RM_polarisation_values}. 

The polarisation map of the central AGN is shown in Fig. \ref{fig:StokesI+P}, panel (c), where only pixels above $6\sigma_{\rm P,i}$ are displayed. Polarised emission is primarily detected from the western lobe, while the eastern lobe appears depolarised, as seen by comparison with the overlaid total intensity contours at 1.5 GHz. The corresponding RM map (Fig. \ref{fig:RM+fracP}, panel (a)) shows a relatively uniform distribution, with values in the source rest frame ranging from $-150 \pm 46$ to $+72 \pm 46 \ \rm rad \ m^{-2}$ and an average around $-115 \pm 32 \ \rm rad \ m^{-2}$. This is consistent with the RM synthesis outputs extracted from the brightest polarised pixel, which reveal a dominant Faraday component at $-14 \pm 5\ \mathrm{rad \ m^{-2}}$ in the observed frame\footnote{The uncertainty of the RM in a single pixel is computed as $\rm FWHM \times (2 \cdot S/N)^{-1}$ in the observed rest frame. In the source frame, the uncertainty is scaled by $(1+z)^2$, with the Galactic RM uncertainty added in quadrature.} (see Fig. \ref{fig:FDF}). After correcting for the Galactic foreground and transforming to the source rest frame, this corresponds to an intrinsic RM of $-127 \pm 46 \ \mathrm{rad \ m^{-2}}$. Additional weaker components are present in the FDF, possibly indicating multiple Faraday components along the LOS (Faraday-complex spectrum). However, their separations are below the resolution limit, and are roughly symmetric and a factor of 25 lower than the main peak ($\approx 4\%$ of its value), consistent with residual contributions from the secondary lobes of the RMTF. Neglecting them does not introduce any significant uncertainty. We note that in the western lobe the RM values are mostly negative, indicating MF components directed away from the observer. The RM modulus decreases towards the core, suggesting a decline in both MF strength and gas density towards the centre. Moving to the eastern lobe, the RM values gradually become positive, suggesting a possible MF reversal along the LOS, although we note that positive RM values are detected over few pixels. The low RM dispersion,\footnote{Computed as the standard deviation of the RM values within a region, and measuring RM fluctuations on pixels of $1.2\arcsec$ ($\sim 10.2$ kpc).} $\sigma_{\rm RM}$, of $36 \pm 11 \ \rm rad \ m^{-2}$ across the source supports the presence of a partially ordered MF. This is further confirmed by the projected MF vectors (Fig. \ref{fig:RM+fracP}, panel (b)), which show coherent alignment across the western lobe.\footnote{The vectors are derived from the polarisation angles, calculated as $\rm \psi = 0.5 \arctan (U,Q) - \mathrm{RM}\cdot \lambda_0^2$ (in degrees), and then rotated by $90^\circ$ to indicate the orientation of the MF vectors.} The fractional polarisation map (Fig. \ref{fig:RM+fracP}, panel (b)) displays values between $(0.30 \pm 0.04)\%$ and $(8.2 \pm 1.2)\%$, showing a clear drop towards the eastern lobe. The highest $\rm F_p$ is found in the outer regions of the western lobe and the average $\rm F_p$ across the source is $(3.8 \pm 1.4)\%$. All measured values are at least three times larger than the maximum leakage of 0.1\%.\footnote{Maximum measured Stokes V/I across the band, extracted from a fixed region at the source position, used to assess the instrumental calibration.} 
\begin{figure}[!htb]
    \vspace{-0.2cm}
    \centering
    \includegraphics[width=0.95\linewidth]{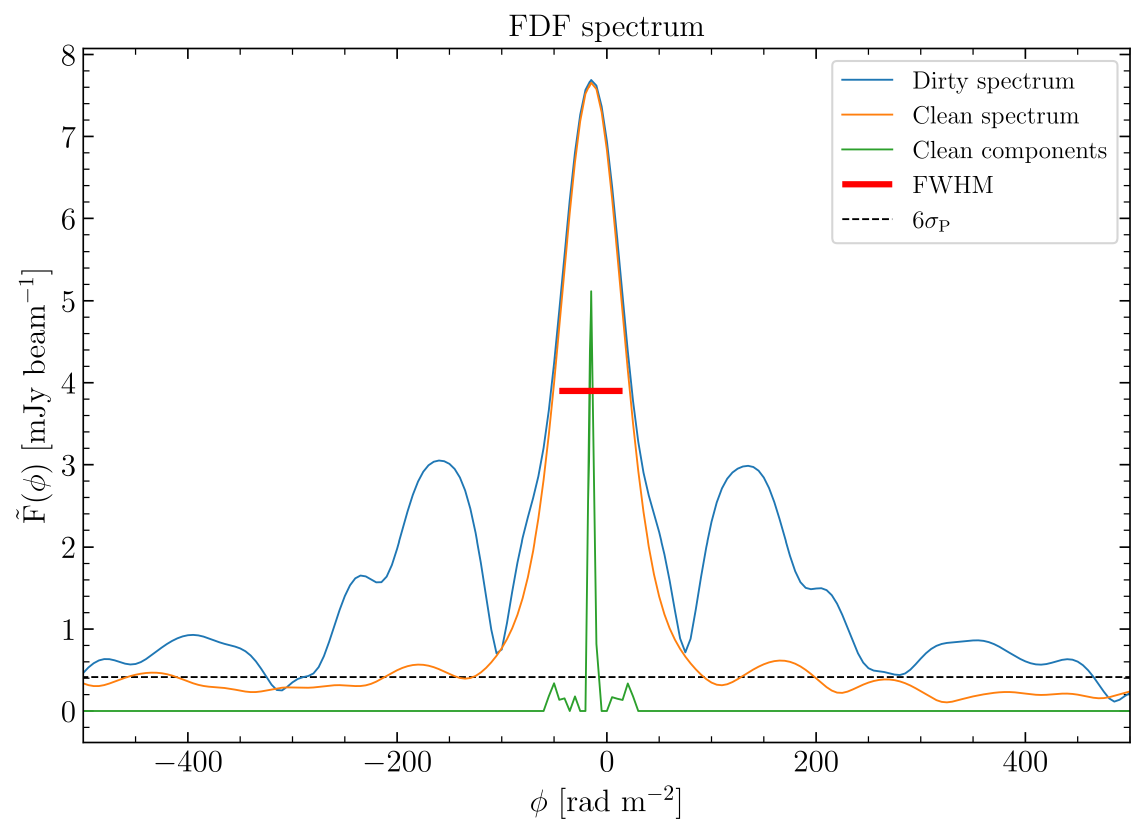}
    \caption{Reconstructed FDF spectrum extracted from the brightest polarised pixel of the central source. The dirty FDF is shown in blue, the CLEANed FDF in orange, and the CLEAN components in green. The FWHM of the RMTF, 60 $\rm rad\ m^{-2}$, is marked in red. The $6\sigma_{\rm P}$ level of the polarised intensity in this pixel is indicated with a black dashed line as a reference. The dominant peak is located at $\phi = -14 \pm 5 \ \rm rad\ m^{-2}$. Values are not corrected to the source rest frame.}
    \label{fig:FDF}
\end{figure}
\begin{table}[!htb]
    \vspace{-0.4cm}
    \centering
    \caption{Rotation measure and polarisation properties of the central AGN.}
    \label{tab:RM_polarisation_values}
    \begin{tabular}{c c c c c}
        \hline
        \noalign{\smallskip}
        \noalign{\smallskip}
          $\langle \mathrm{RM} \rangle \pm \mathrm{err_{\langle RM \rangle}}$ & & $\mathrm{\sigma_{RM}} \pm \mathrm{err_{\sigma_{RM}}}$ & & $\mathrm{F_p} \pm \mathrm{err_{F_p}}$ \\
          $\mathrm{[rad \ m^{-2}]}$ & & $\mathrm{[rad \ m^{-2}]}$ & & $[\%]$ \\
         (1) & & (2) & & (3) \\
         \noalign{\smallskip}
         \hline
         \noalign{\smallskip}
         \noalign{\smallskip}
         $-115 \pm 32$ & & 36 $\pm$ 11 & & 3.8 $\pm$ 1.4 \\ 
         \noalign{\smallskip}
         \hline
    \end{tabular}
    \tablefoot{(1): Average RM with associated uncertainty; (2): RM dispersion with associated uncertainty; (3): Fractional polarisation with associated uncertainty. All the values are in the source rest frame. The uncertainties in $\langle \mathrm{RM} \rangle$ and in $\sigma_{\mathrm{RM}}$ are calculated as $\mathrm{err_{\langle RM \rangle}}=\sigma_{\mathrm{RM}}\cdot(\rm n_{\mathrm{beam}})^{-1/2}$ and $\mathrm{err_{\sigma_{RM}}}=\sigma_{\mathrm{RM}} \cdot (2 \rm n_{\mathrm{beam}})^{-1/2}$, respectively, where $\rm n_{\mathrm{beam}}$ refers to the number of independent beams over which the RM is calculated (here 5.2). For the average RM, we account for the redshift-corrected uncertainty in the Galactic RM by summing it in quadrature. The uncertainty in $\mathrm{F_p}$ is based on the propagation of the uncertainties as $\mathrm{err_{F_p}}=\mathrm{F_p}/2\big[\left(\sigma_{\mathrm{P}}/{\mathrm{P}}\right)^2+\left(\sigma_{\mathrm{I}}/\mathrm{I}\right)^2\big]^{1/2}$ (here we assume that the I, Q, and U parameters exhibit no cross-covariance, and that $\rm \sigma_Q = \sigma_U = \sigma_P/2$).}
    \vspace{-0.5cm}
\end{table}
\section{QU fitting results} \label{QU fitting results}
For the QU fitting, several polarisation models were tested (see Appendix \ref{QU fitting-2}); however, none of them perfectly fits the data ($16 \lesssim \chi^2_\nu \lesssim 62$). In this section, we present the results of model n3, which yields a reduced chi-squared of $\chi^2_\nu = 14.7$. Although this value remains high, this model reproduces the main features of the observed Stokes Q and U spectra better than the other tested configurations. The fitting is performed over an integrated region defined by total intensity ($\rm I \ge 3\sigma_{\rm I}$), thus encompassing the emission from both the lobes, even though polarised emission is detected only from the western one. This model includes two components: one Faraday-simple (which we attribute to a foreground magnetised medium, i.e. the RM synthesis peak), and one with differential Faraday rotation (where the synchrotron-emitting plasma of the radio lobe is mixed with a thermal gas component). Both are subject to external depolarisation from a foreground turbulent screen, which we identify as the proto-ICM (see Eq. n3 in Appendix \ref{QU fitting model}). The best fit model among those tested is shown in Fig. \ref{fig:QUfit-n3}, while the corresponding corner plot is provided in Appendix \ref{QUfitting plots} (Fig. \ref{fig:CP-n3}).
\begin{figure}[!htb]
    \centering
    \includegraphics[width=1\linewidth]{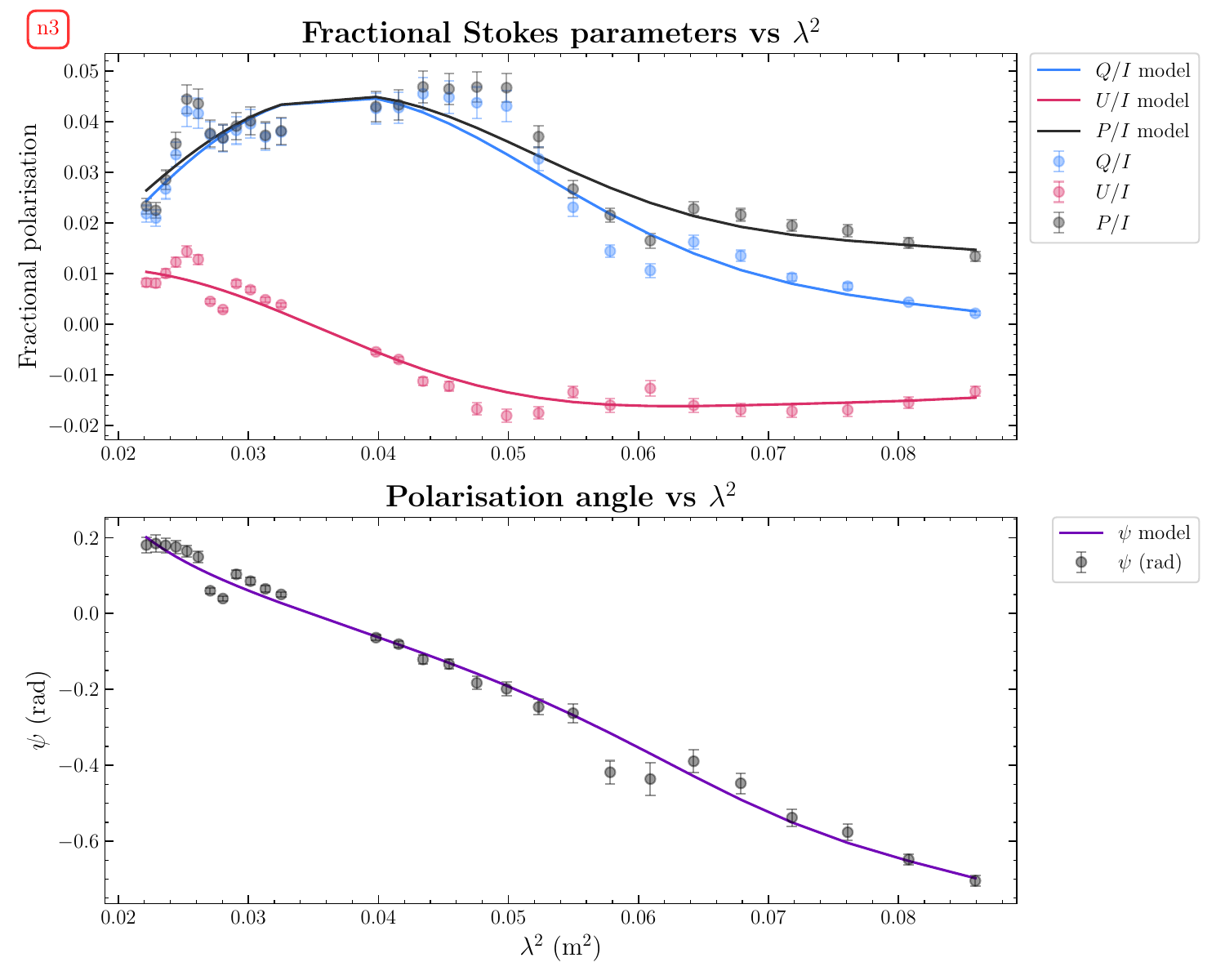}  
    \caption{Best fit of the fractional polarisation (P/I, Q/I, and U/I) and polarisation angles as a function of $\lambda^2$ for model n3. The data points represent the observations; the straight lines are the model predictions (P/I: black, Q/I: blue, U/I: red, $\psi$: purple).}
    \label{fig:QUfit-n3}
\end{figure}
\\
The RM of the simple foreground component is $-13.5 \pm 0.4 \ \mathrm{rad \ m^{-2}}$ ($-123.6 \pm 27.5\ \mathrm{rad \ m^{-2}}$ rest frame), consistent with the observed average RM. The thick lobe component has a mean RM of $-61.7 \pm 3.8\ \mathrm{rad \ m^{-2}}$ ($-480.3 \pm 39.2\ \mathrm{rad \ m^{-2}}$ rest frame) and an internal dispersion of $108.8 \pm 1.1 \ \mathrm{rad \ m^{-2}}$ ($805.1 \pm  8.1\ \mathrm{rad \ m^{-2}}$ rest frame), while the common external dispersion is $8.6 \pm 0.1 \ \mathrm{rad \ m^{-2}}$ ($63.6 \pm 0.7 \ \mathrm{rad \ m^{-2}}$ rest frame). Although this external dispersion exceeds the observed one (see Tab. \ref{tab:RM_polarisation_values}), the fitting also includes the depolarised eastern lobe, which may enhance the fitted external dispersion. Nevertheless, this is not sufficient to entirely depolarise the western lobe. The thick component is not detected in the FDF because its large internal Faraday dispersion makes it broader than the maximum recoverable Faraday scale of the data ($\Delta\phi_{\max}=143 \ \mathrm{rad\ m^{-2}}$), and the combined effect with the wavelength-dependent depolarisation reduces its polarised signal. Consequently, it does not form a detectable peak in the FDF and only the thin RM component is observed.

These results complement the RM synthesis by suggesting that multiple Faraday components are required to reproduce the observed Stokes Q and U spectra, together with both internal depolarisation associated with the radio plasma and external depolarisation arising from a foreground turbulent screen. In particular, the QU fitting favours the presence of an internal Faraday-thick component, which we interpret as magnetised relativistic plasma associated with the AGN (e.g. the radio lobe), and mixed with the surrounding thermal gas. This suggests that the AGN may contribute not only to the synchrotron emission, but also to the magnetisation of the environment. Overall, the parameters found by fitting model n3 support a scenario in which both a foreground magnetised gas and the lobe itself contribute to the Faraday rotation, and additional depolarisation may arise from an external turbulent screen (e.g. the proto-ICM). However, model n3 yields a high $\chi^2_\nu$ value, showing that simple analytical models are not able to fully capture this complex environment. Therefore, we still rely primarily on the RM synthesis results, while the QU fitting provides supporting evidence that the AGN may influence the proto-ICM properties.

\section{Constraints on the MF} \label{MF estimate}
Estimating the MF strength along the LOS is not straightforward, as it depends on the unknown 3D structure of the field and the physical properties of the Faraday rotating region. In this section, we first interpret the results presented in the previous section by describing the configuration and properties of the Faraday rotating medium (Sect. \ref{Medium Properties}). We then followed a three-step approach to constrain the MF in CARLAJ1510. First, we used the RM formula (Eq. \ref{eq:RM}) to provide a first-order estimate of the LOS MF strength in front of the polarised lobe (Sect. \ref{RM formula}). Then, we ran semi-analytical 3D simulations of a random MF and created mock RM maps, which we compared with the observed RM values (Sect. \ref{MF simulations}). Finally, we used the non-detection of the polarised radiation from the eastern lobe to place lower limits on the MF strength in the proto-ICM (Sect. \ref{LL}).
\vspace{-0.2cm}
\subsection{Properties of the Faraday rotating medium} \label{Medium Properties} 
Assuming both lobes are intrinsically polarised at the same level and inclined with respect to the LOS, the observed asymmetry is likely due to external depolarisation from a turbulent Faraday rotating screen along the LOS  \citep{Burn_1966, Tribble_1991}, the so-called Laing-Garrington effect \citep{Laing_1988, Garrington_1988, Tribble_1992}. In this framework, the receding (eastern) lobe lies behind a denser, more turbulent magneto‑ionic medium with a higher $\sigma_{\rm RM}$, causing stronger depolarisation, while the approaching (western) lobe is viewed through a shorter path, preserving more of its polarised radiation. The homogeneous RM of the polarised lobe further indicates the presence of an ordered foreground external screen. These features can be explained by the lobe expansion into the ambient medium, but differ in where the MF ordering occurs. In particular, we identified two potential scenarios. In the first one, ram pressure compresses the plasma and the MF within the outer lobe, producing a uniform LOS field and low RM dispersion, consistent with the large‑scale alignment of the polarisation vectors. In the second one, shear at the lobe boundary stretches the ambient field in a thin external magnetised sheath, creating an ordered Faraday screen that also yields low RM dispersion and aligned vectors. Thus, the observed pattern may reflect MF ordering either within the lobe, or in a surrounding sheath (see Fig. \ref{fig:Sketch}). \citet{Guidetti_2011} reported a similar condition using high-resolution, multifrequency JVLA observations, and finding ‘banded’ RM patterns around nearby radio galaxies, interpreted as magnetic draping and compressed shells of thermal plasma formed as expanding lobes stretch the ambient MF. We note that we cannot exclude a more complex scenario in which both the cases occur simultaneously, although this possibility is not treated here. In any case, the western lobe’s polarisation suggests that the surrounding medium is already magnetised, either by the AGN itself or by previous processes. 

The intrinsic orientation of the AGN and the gas distribution along the LOS are uncertain. Although the inclination is not directly constrained due to unresolved jet emission, an angle range of $30^{\circ}-45^{\circ}$ can explain the observed Laing-Garrington effect, as it provides enough path length difference through the external medium to depolarise the eastern lobe while leaving the western lobe polarised. Smaller angles ($<30^{\circ}$) would imply a more beamed, blazar-like core, whereas larger angles ($>45^{\circ}$) would not produce the observed polarisation asymmetry.

For the structure of the external medium, we considered two limiting scenarios: (1) a homogeneous environment in which the AGN is fully embedded, so depolarisation arises from different path lengths through the gas (Fig. \ref{fig:Sketch}, left); and (2) an inhomogeneous environment with gas concentrated around the eastern lobe (higher electron density), while the western lobe lies behind a compressed Faraday screen (Fig. \ref{fig:Sketch}, right). On the other hand, we cannot determine whether a similar screen affects the eastern lobe, as it is depolarised.

\begin{figure}[!htb]
    \centering
    \includegraphics[width=1\linewidth]{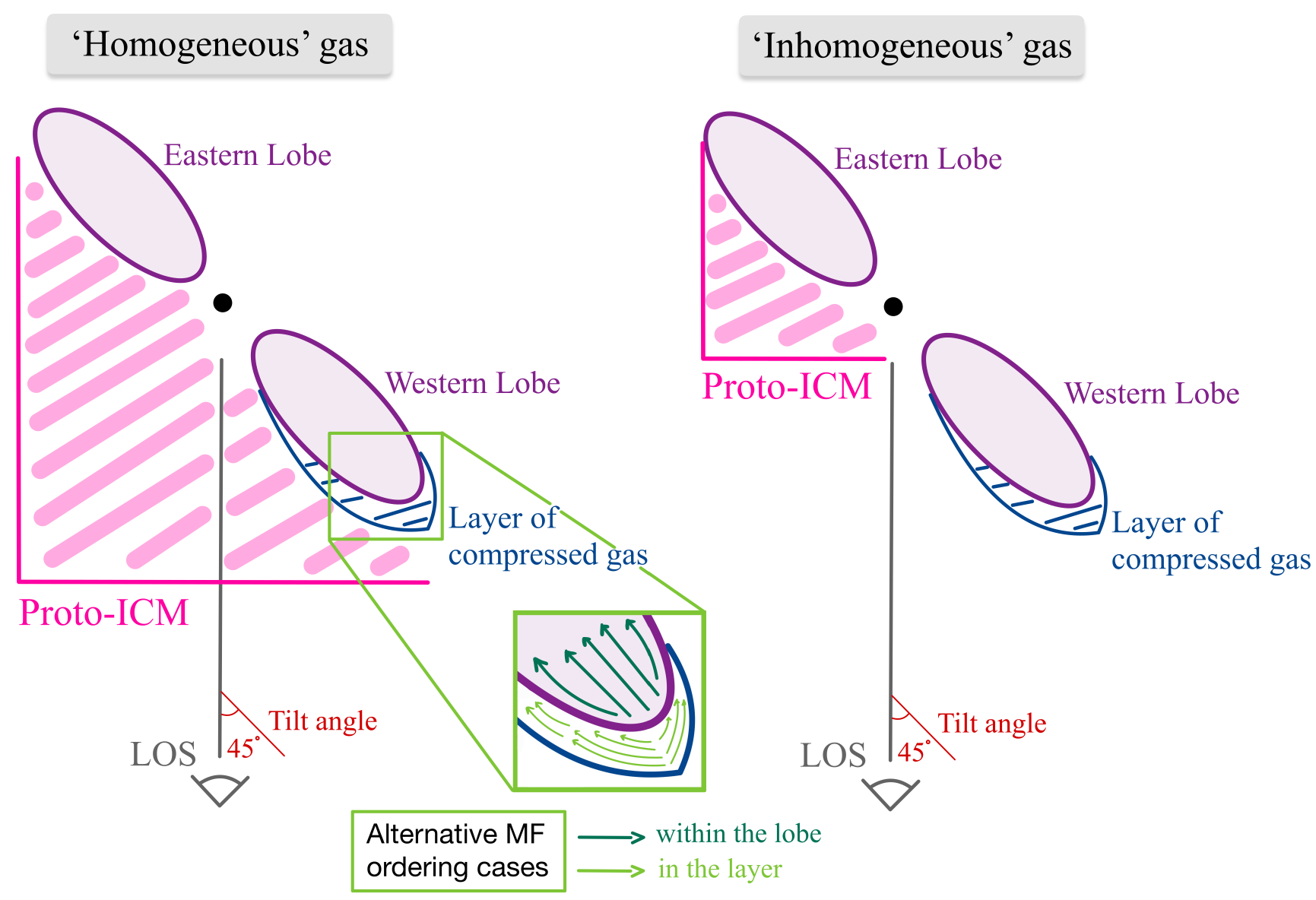}
    \caption{Schematic representation of the two limiting scenarios for the distribution of the external magnetised medium around the AGN ($45^\circ$ with respect to the LOS) and the MF ordering. Left: Homogeneous case. The AGN is fully embedded in the proto-ICM, resulting in different path lengths through the magnetised gas for the two lobes. Right: Inhomogeneous case. The medium is asymmetric, with the proto-ICM located preferentially in front of the eastern lobe. In both cases, the western lobe may lie behind a thinner compressed layer of gas. The MF ordering may occur either within the lobe (dark green arrows) or in the external compressed sheath (light-green arrows). The sketch is not to scale and shows a top-down view. The proto-ICM gas distribution is assumed to be symmetric around the AGN; only the foreground is shown for simplicity.}
    \label{fig:Sketch}
    \vspace{-0.2cm}
\end{figure}
Since we lack direct constraints on the gas density distribution from X-ray or SZ observations, to estimate the gas density of the proto-ICM we used $n_e$ values inferred by cosmological simulations. In particular, we extracted the spherically averaged $n_e(r)$ profiles for all proto-clusters with $\rm M_{500c} > 10^{13} \ M_\odot$ in the Magneticum Box2b/hr simulation and stacked them (a more complete description will be covered in \citealt{VallesPerez_2026}). We then computed the electron column densities by integrating these profiles along the same LOS depth, and averaged them in the projected area of diameter $\approx 250 \ \mathrm{kpc}$ matching the lobes extension. Finally, we selected the constant electron densities as those that would reproduce, under identical LOS integration and aperture on the plane of the sky, the same averaged column density as the stacked profiles. The corresponding $n_e$ values, which depend on the assumed LOS integration length, are reported in the following subsections for the different cases considered. We note that, in principle, radial density profiles would provide a more realistic description than a constant density. However, there are no direct observational constraints on the gas density, and calibrated $n_e(r)$ profiles covering the full radial range required for our analysis were still under preparation during our modelling phase (\citealt{VallesPerez_2026}). Adopting a constant electron density is a conservative choice, as it avoids introducing additional assumptions on the radial gas distribution while still accounting to the same contribution to the RM. In a follow-up study, we will use density profiles to more accurately account for the RM contributions from dense regions.
\vspace{-0.3cm}
\subsection{MF strength from the RM formula} \label{RM formula}
To obtain a first-order estimate of the MF strength along the LOS responsible for the observed RM in the western lobe, we inverted the RM equation in the source rest frame (Eq. \ref{eq:RM}), following the approach described by \citet{Anderson_2022}. This method relies on the simplifying assumption that both the thermal electron density and the MF are uniformly distributed throughout the Faraday rotating medium. We adopted an integration length of 100 kpc, which is comparable to the characteristic extent of the proto-ICM inferred for the Spiderweb proto-cluster, and assumed a uniform gas density of $\rm n_e \sim 10^{-2} \ \rm cm^{-3}$ based on simulations (Sect. \ref{Medium Properties}). This density agrees with the value measured within 100 kpc in the Spiderweb system ($1.5 \times 10^{-2}\ \rm cm^{-3}$; \citealt{Tozzi_2022b}; \citealt{Anderson_2022}), once rescaled to the redshift of CARLAJ1510 using the expected redshift evolution $\rm n_e(z) = n_{e,0}(1+z)^3$. Using the average observed RM of $-115 \pm 32 \ \rm rad \ m^{-2}$ (Tab. \ref{tab:RM_polarisation_values}), we derived an MF strength of $(0.14 \pm 0.04) \ \mu\rm G$, corresponding to a comoving field of $(0.019 \pm 0.005) \ \mu\rm G$.
\vspace{-0.3cm}
\subsection{Comparison with semi-analytical simulations} \label{MF simulations}
As a second step, we investigated whether the observed RM values can be explained by a random MF or if they indicate the presence of a coherent component, as discussed in Sect. \ref{Medium Properties}. We further examined the spatial scales of the MF fluctuations and assessed whether the Faraday rotation originates within 100 kpc in front of the polarised western lobe. 

Previous radio observations (e.g. \citealt{Murgia_2004}; \citealt{Bonafede_2010}) and cosmological simulations (e.g. \citealt{Vazza_2018}; \citealt{DF_2019}) show that MFs in galaxy clusters typically fluctuate across a range of spatial scales and that the MF energy scales with the gas energy density. This complexity implies that a simple uniform MF model is insufficient, motivating the use of 3D simulations to capture the more realistic nature of the MF and its effect on the RM distribution. To this end, we used the \texttt{MIRO'} code (\citealt{Bonafede_2013}), which generates 3D models of the random MF and the gas density distribution and computes the corresponding 2D RM maps by solving Eq. \ref{eq:RM}. These synthetic maps are then compared with the observed RM map to infer the MF properties responsible for the Faraday rotation. For a detailed description of the simulations methodology, we refer to \citet{Bonafede_2013}. Here, we provide a brief summary specific to our case:
\begin{itemize}
    \itemsep3pt
    \item[-] The simulation takes as input both the size of the simulated box and the resolution of the individual cells. To match the observations, we adopted an angular resolution of $0.6^{\prime\prime}$ per pixel (i.e. 5.1 kpc at the proto-cluster redshift), and generated a cube of $400^3$ pixels (i.e. $\sim 2^3 \rm \ Mpc^3$).
    \item[-] The thermal gas distribution was modelled as a 3D cube with a constant mean physical electron density of $10^{-2}\ \mathrm{cm}^{-3}$ (see motivation in the previous section), with 5\% Gaussian random fluctuations in the gas density of the cells. This choice ensures a conservative estimate of the MF, as it gives more weight to MF fluctuations in the RM dispersion.
    \item[-] The MF was generated as a 3D model assuming an analytical Kolmogorov power-law spectrum, $\rm E_{B}(k) \propto k^{-(n+2)} = k^{-11/3}$, where $\rm k = (\sum_i k_i^2)^{1/2}$ (with $i=1,2,3$) is the wavenumber corresponding to the physical scale of the MF fluctuations (e.g. $\rm \Lambda \propto 1/k$), and $\rm n=5/3$ is the spectral index.
    \item[-] The maximum fluctuating scale of the MF components in Fourier space was defined as $\rm k_{\rm max} = (400 \cdot 0.5) - 1 = 199$, which gives $\rm \Lambda_{\rm min} = (400 \cdot 5.1) \cdot k_{\rm max}^{-1} = 10.3\ \rm kpc$.
    \item[-] The MF strength was assumed to scale with the thermal gas density as $\rm |\mathbf{B}(\mathbf{r})| \propto n_e(\mathbf{r})^{\eta}$, with $\eta = 0.5$ \citep{Bonafede_2010}.
\end{itemize}

Overall, the MF model depends on five key parameters: the minimum and maximum fluctuation scales, $\Lambda_{\rm min}$ and $\Lambda_{\rm max}$, the radial scaling index, $\eta$, the MF power spectrum index, n, and the physical MF normalisation, $\rm B_{\rm norm}$ (i.e. the MF averaged over the box volume). A summary of the adopted (fixed) parameter values is reported in Table \ref{tab:Sim specifics}.
\begin{table}[!htb]
    \caption{List of fixed parameters adopted for the simulations.}
    \begin{center}
        \begin{tabular}{ c c c c c c c} 
            \hline
            \noalign{\smallskip}
            \noalign{\smallskip}
            Size of the box & & & 400 pxl \\ 
            \noalign{\smallskip}
            Resolution & & & 5.1 kpc \\
            \noalign{\smallskip}
            $\rm \Lambda_{min}$ & & & 10.3 kpc \\ 
            \noalign{\smallskip}
            $\eta$ & & & 0.5 \\
            \noalign{\smallskip}
            n & & & 5/3 \\
            \noalign{\smallskip}
            \hline
     \end{tabular}
     \label{tab:Sim specifics}
     \end{center}
     \vspace{-0.4cm}
\end{table}   
\\
The two free parameters, instead, are $\Lambda_{\rm max}$ and $\rm B_{\rm norm}$, and can be investigated through the comparison with our observations. We tested two MF power spectra, with $\rm \Lambda_{max} = [51, 102]$ kpc, and several normalisations in the range $\rm B_{norm}=[0.5,...,3.0] \ \mu$G. We integrated the simulation cubes along the LOS from the centre of the box out to 100 kpc. The resulting RM maps were then convolved with a Gaussian beam of $\sim 59$ kpc, and re-binned to a resolution of $1.2^{\prime\prime}$ (corresponding to 10.3 kpc) and to a final size of $2^2$ Mpc$^2$. In the end, the maps were masked following the same procedure applied to the observed RM data (see Sect. \ref{Final maps}).

For each model, we ran three simulations and, in general, none of the tested combinations of parameters gives a model able to reproduce the observed average RM and RM dispersion of the western lobe (see Tab. \ref{tab:Randomsimvalues}). This indicates that the Faraday rotation is not primarily caused by a turbulent medium with a fully random MF configuration, but rather by a medium with a well-ordered MF structure on the scale of the radio lobe, as indicated by the RM results (see Sect. \ref{RMS results}). However, the presence of a turbulent component in the proto-cluster cannot be ruled out, as it is suggested by the depolarisation of the eastern lobe and by the QU fitting results (Sect. \ref{QU fitting results}).
\begin{table}[!htb]
    \caption{$\rm \langle RM \rangle$ and $\rm \sigma_{RM}$ extracted from the simulations.}
    \begin{center}
        \begin{tabular}{ c c c c c c c} 
            \hline
            \noalign{\smallskip}
            \noalign{\smallskip}
            $\rm \Lambda_{max}$ & & $\rm B_{norm}$ & &  $\langle \mathrm{RM} \rangle \pm \mathrm{err_{\langle RM \rangle}}$ & & $\mathrm{\sigma_{RM}} \pm \mathrm{err_{\sigma_{RM}}}$ \\
            $\rm [kpc]$ & & $[\mu$G] & & $\mathrm{[rad \ m^{-2}]}$ & & $\mathrm{[rad \ m^{-2}]}$ \\
            (1) & & (2) & & (3) & & (4)\\
            \noalign{\smallskip}
            \hline
            \noalign{\smallskip}
            \noalign{\smallskip}
            51 & & 0.5 & & $0.6 \pm 0.8 $ & & $2.5 \pm 0.5$ \\[4pt]
            102 & & 0.5 & & $1.9 \pm 1.7 $ & & $20.6 \pm 3.1$ \\[4pt]
            51 & & 1.0 & & $-0.5 \pm 2.8 $ & & $4.7 \pm 0.6$ \\[4pt]
            102 & & 1.0 & & $-1.5 \pm 4.7 $ & & $37.4 \pm 7.0$ \\[4pt]
            51 & & 1.5 & & $-1.2 \pm 4.1 $ & & $6.3 \pm 1.1$ \\[4pt]
            102 & & 1.5 & & $15.9 \pm 9.4 $ & & $49.2 \pm 8.8$ \\[4pt]
            51 & & 2.0 & & $8.5 \pm 7.4 $ & & $8.2 \pm 0.4$ \\[4pt]
            102 & & 2.0 & & $-3.0 \pm 8.6 $ & & $71.1 \pm 8.8$ \\[4pt]
            51 & & 3.0 & & $1.2 \pm 6.0 $ & & $11.5 \pm 1.5$ \\[4pt]
            102 & & 3.0 & & $4.1 \pm 10.1 $ & & $105.2 \pm 10.5$ \\
            \noalign{\smallskip}
            \hline
     \end{tabular}
     \tablefoot{(1): Maximum spatial fluctuation of the MF; (2): MF normalisation; (3): Average RM, computed as the mean over three simulations (uncertainty: standard deviation); (4): RM dispersion, computed as the mean over three simulations (uncertainty: standard deviation).}
     \label{tab:Randomsimvalues}
     \end{center}
     \vspace{-0.6cm}
\end{table}
\vspace{-0.1cm}
\subsection{Lower limits on the MF strength from the non-detection} \label{LL}
Assuming that the depolarisation of the eastern lobe is external and caused by a randomly magnetised proto-ICM along the LOS, we used the non-detection of polarised emission from the lobe to derive a lower limit on the $\sigma_{\mathrm{RM}}$ required to explain the observed depolarisation, and consequently to place a lower limit on the physical MF strength. The upper limit on the fractional polarisation of the depolarised lobe was estimated by masking the Stokes I map (Fig. \ref{fig:StokesI+P}, panel (a)) and the polarisation noise map below $6\sigma_{\rm P,i}$ and above $\rm 3\sigma_I$. The selected pixels are contained in a region on the eastern side, delimited by the host galaxy position and the $3\sigma_{\rm I}$ contour (see Fig. \ref{fig:MockRM} in Appendix \ref{MockRM}). We then computed the ratio between six times the average noise in the masked polarisation map and the average total surface brightness only in the eastern lobe region, resulting in an upper limit of $\sim2\%$ for the fractional polarisation. We adopted a conservative approach by using the most polarised pixel on the western side and the outermost region on the eastern side, in order to minimise contamination between the two regions. Since the eastern and western lobes are not well separated, the western lobe can contribute to the total intensity in the eastern region, which would slightly increase $I$ and therefore reduce the estimated $\rm F_p$. As a result, our 2\% upper limit is conservative.

Making the reasonable assumption that the two lobes are identical in the structure and that the eastern lobe is intrinsically polarised similarly to the western lobe (up to $\sim8\%$), the $\sigma_{\mathrm{RM}}$ required at 1.5 GHz to reduce the fractional polarisation to $\sim 2\%$ can be estimated using the \citet{Burn_1966} depolarisation model
\begin{equation}
    \Pi(\lambda) = \Pi_0 \, e^{-2\sigma_{\rm RM}^2 \lambda^4},
\end{equation}
where $\Pi(\lambda)$ is the observed fractional polarisation, $\Pi_0$ is the intrinsic polarisation, and $\lambda$ is the observing wavelength. Using $\Pi(\lambda) \sim 2\%$, $\Pi_0 \sim 8\%$, and $\lambda = 0.2$ m (corresponding to 1.5 GHz), we found that the minimum $\sigma_{\mathrm{RM}}$ required is $\sim21\ \rm rad\ m^{-2}$. Hence, in front of the eastern lobe we can put a rest frame lower limit of $\rm \sigma_{RM}> 21 \cdot (1+z)^2=155\ \rm rad\ m^{-2}$.\\
\\
To investigate the range of MF strengths that could explain the observed depolarisation, we generated synthetic RM maps using the simulation parameters described in Table \ref{tab:Sim specifics} and varying the integration length along the LOS. Assuming the optical position of the host galaxy as the reference point separating the two lobes (although some displacement could be present), the projected size of the western lobe is approximately 170 kpc. Assuming that both lobes extend equally in three dimensions and that the source is inclined at an angle of $45^\circ$ with respect to the LOS, this corresponds to a LOS depth of $340$ kpc, in the case of uniform gas distribution (homogeneous gas; see Fig. \ref{fig:Sketch}). We initially tested this scenario, and then reduced the LOS length to 170 kpc in the case where only the eastern lobe is embedded in the depolarising medium (inhomogeneous gas; see Fig. \ref{fig:Sketch}). Additionally, we tested a LOS depth of 600 kpc, corresponding to a $30^\circ$ inclination from the LOS. Regarding the gas density, we adopted the following average physical electron densities obtained from simulations (see Sect. \ref{Medium Properties}): $1.052 \times 10^{-2} \ \rm cm ^{-3}$ (170 kpc), $6.973 \times 10^{-3} \ \rm cm ^{-3}$ (340 kpc), $4.716 \times 10^{-3} \ \rm cm ^{-3}$ (600 kpc). We tested the set of free parameters, $\rm \Lambda_{max} = [51, 102, 255, 510]$ kpc and $\rm B_{norm}=[0.5,...,5.0] \ \mu$G.

The synthetic RM maps were processed following the same procedure described in the previous section; however, in this case, the mask applied is identical to that used for deriving the upper limit on the fractional polarisation. To determine the lower limit of the MF strength (physical and comoving), we performed three simulations for each model. We then compared the mean RM dispersion extracted from the eastern lobe region (with the standard deviation used as scatter) with the observational lower limit of $155 \ \rm rad \ m^{-2}$.
\vspace{-0.1cm}
\begin{figure}[!htb]
    \centering
    \includegraphics[width=1\linewidth]{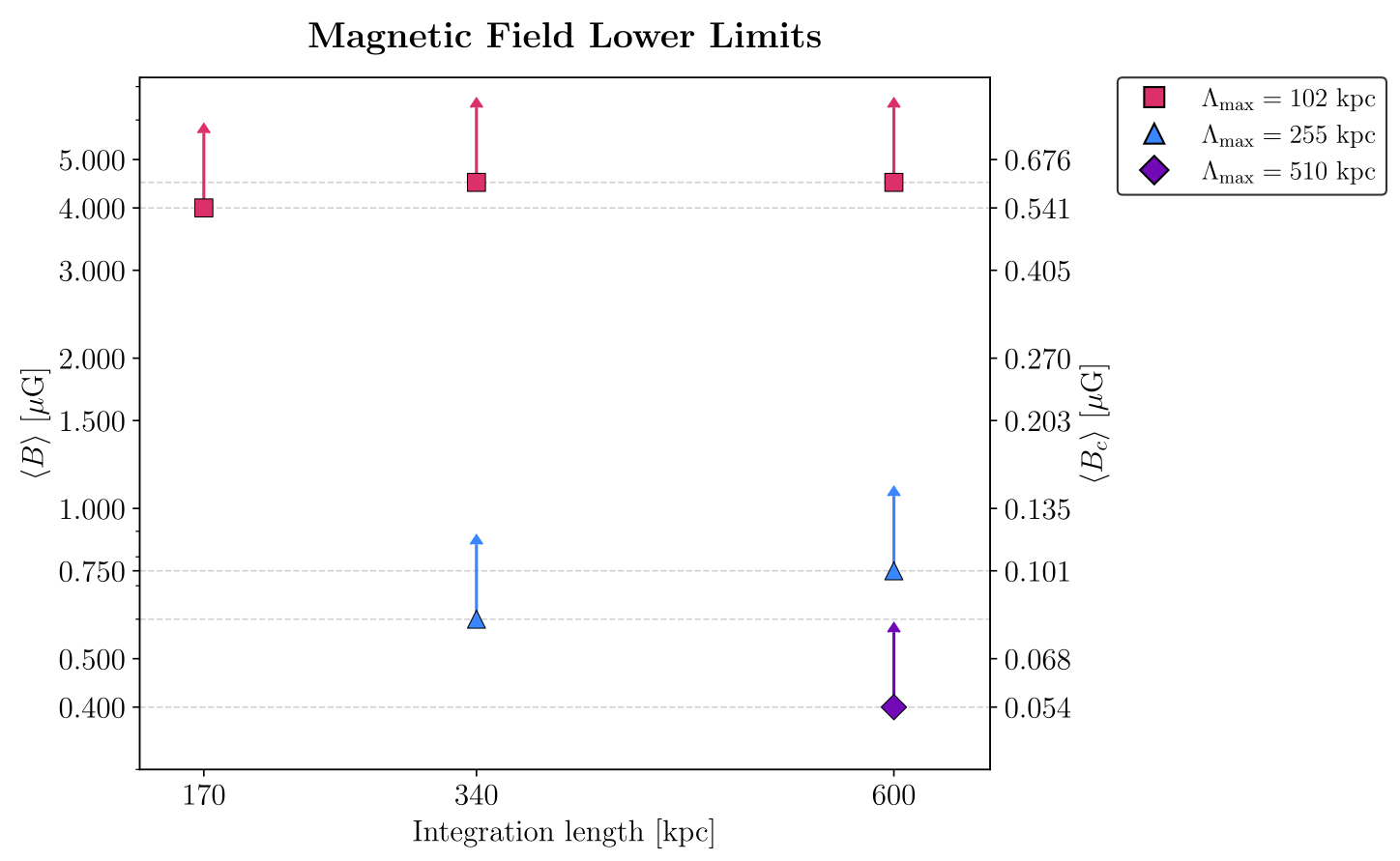}
    \caption{Lower limits on the average MF strength as a function of LOS integration length and maximum fluctuation scale of the MF. From top to bottom ($\rm \Lambda_{max}$): magenta square (102 kpc), light blue triangle (255 kpc), and purple diamond (510 kpc). The right-hand y-axis shows the corresponding comoving MF strength ⟨$\rm B_c$⟩. The y-axes are in logarithmic scale for a better visualisation of the data. }
    \vspace{-0.3cm}
    \label{fig:MFLL}
\end{figure}

Three representative examples of mock RM maps are reported in Appendix \ref{MockRM} (Fig. \ref{fig:MockRM}). The results, instead, are shown in Fig. \ref{fig:MFLL}, which displays the minimum physical and comoving MF strengths required to reproduce the observed level of depolarisation, as a function of the LOS integration length and maximum fluctuating spatial scale. Depending on the scales of the spectrum, the values range from a few microgauss (for small-scale turbulence) to $0.4 \ \mu$G (for larger scales), and decrease with longer integration lengths. In case of $\rm \Lambda_{max} = 51$ kpc, the required average physical MF would exceed 10 $\mu$G. From that, we can place a non-zero lower limit of 0.4 $\mu$G on the physical MF ($0.05 \ \mu$G on the comoving one). To our knowledge, this is the first constraint indicating non-zero MFs in proto-clusters.

\section{Discussion} \label{Discussion}
In this section, we compare our results with those obtained for the Spiderweb proto-cluster (Sect. \ref{Comparison Spiderweb}). In Sect. \ref{Comparison literature} we place our findings in the context of the existing literature, whereas in Sect. \ref{Other scenarios} we explore possible physical interpretations for the observed polarisation and depolarisation. Finally, in Sect. \ref{Limitations} we discuss the main limitations of our observations and MF modelling, and outline potential directions for future work.
\vspace{-0.2cm}
\subsection{Comparison with the Spiderweb proto-cluster} \label{Comparison Spiderweb}
We compare our results with those of \citet{Anderson_2022}, who studied the Spiderweb proto-cluster at $z = 2.16$. Their analysis revealed complex Faraday structures with sharp RM gradients ranging from $-4400$ to $+5800\ \rm rad\ m^{-2}$ (in the source frame), occurring on scales smaller than the synthesised beam. These strong RM variations suggest rapid changes in the LOS MF direction, consistent with the presence of helical or toroidal fields near the jets. In contrast, our RM map shows a more uniform distribution, with values between $-150$ and $+72\ \rm rad\ m^{-2}$, a low RM dispersion ($36\ \rm rad\ m^{-2}$), and a coherent alignment of projected MF vectors. Although we cannot exclude an RM originating within the lobes (see Sect. \ref{QU fitting results}), we interpret this configuration as the RM produced by outer regions of the lobe compressed by expansion or the result of compression of a thin, magnetised foreground medium.

The RM values observed in our study are roughly one order of magnitude lower than those in \citet{Anderson_2022}, which is expected given the different frequency setups. Their high-frequency observations are sensitive to very large Faraday depths (up to $\sim 1.4 \times 10^6\ \rm rad\ m^{-2}$), but have coarser Faraday resolution. In contrast, our lower-frequency data provide a narrower FWHM ($60\ \rm rad\ m^{-2}$) and are sensitive up to $\sim 1300\ \rm rad\ m^{-2}$, making them well suited for detecting weaker Faraday rotation from the diffuse proto-ICM. Consistently, we do not detect the extreme RM values associated with AGN jets of the Spiderweb radio galaxy, likely because higher-frequency observations would be required to constrain the MF in the jets. We also find a notable difference in the fractional polarisation. In our observations, the western lobe shows 3.8\% polarisation, while the eastern lobe is depolarised. By comparison, \citet{Anderson_2022} measured much higher fractional polarisations at higher frequencies (25–40\% in the western jet, 10–25\% in the eastern hotspot). This discrepancy can be explained by frequency-dependent and beam depolarisation effects. The higher-frequency data of \citet{Anderson_2022}, with a resolution of $0.5^{\prime\prime}$ ($\approx$ 4 kpc), can resolve compact regions with high RM close to the jets. In contrast, our lower-frequency observations, with a larger beam of $7^{\prime\prime}$ ($\sim$ 59 kpc), average the signal over a much wider area and are therefore mainly sensitive to the extended, weakly magnetised proto-ICM.\\
\\
To understand the origin of the RM observed in the Spiderweb proto-cluster, \citet{Anderson_2022} adopted a simple analytical model (their Eq. 4) and calculated the expected $\sigma_{\rm RM}$ for various combinations of MF strength and coherence length. They found that a MF of $\sim 9\ \mu$G in the interaction region ($\sim 10$ kpc) would require an unrealistically large MF reversal scale, supporting a localised origin for the RM, likely in the immediate vicinity of the AGN. In our study, using the RM equation (Eq. \ref{eq:RM}) and assuming a uniform electron density and constant field along the LOS, we derive a MF strength of $1.4 \pm 0.4\ \mu$G for a typical gas density of $\rm n_e \sim 10^{-2}\ \rm cm^{-3}$ and a path length of 10 kpc. This value is significantly lower than the higher field strengths they inferred for the interaction ($\sim 10 \ \mu$G), jet ($\sim 35 \ \mu$G) and hotspot regions ($\sim 90 \ \mu$G). This supports the interpretation that in CARLAJ1510 we are tracing a different, less magnetised component of the environment, possibly associated to the external region of the lobe or a compressed surrounding gas.

Conversely, \citealt{Anderson_2022} assumed that low RM values could result from either weak MFs or frequent field reversals along the LOS. For instance, a field of $10\ \mu$G with a reversal scale of $0.1$ kpc, or a field of $1\ \mu$G with a $10$ kpc reversal scale, would produce similar RM values once integrating for 100 kpc along the LOS. Based on this analysis, they placed an upper limit of $1\ \mu$G on the physical MF strength in the proto-ICM, that translates to a comoving MF of $0.1\ \mu$G. Following the proposed configuration (i.e. adopting different LOS paths) and MF auto-correlation lengths, we constrain in CARLAJ1510 a lower limit of $0.05 \ \rm \mu$G in the comoving MF. This is consistent with what derived for the Spiderweb, suggesting that weak but non-negligible MFs are already in place in the proto-ICM at high redshift.
\vspace{-0.2cm}
\subsection{Implications for early ICM magnetisation} \label{Comparison literature}
In nearby galaxy clusters, polarisation studies reveal high RM values, ranging from hundreds to thousands of rad m$^{-2}$, with multiple MF reversals along the LOS on scales smaller than or comparable to the radio sources (e.g. \citealt{Bonafede_2010}; \citealt{Govoni_2010}, \citeyear{Govoni_2017}; \citealt{Boehringer_2016}; \citealt{Pagliotta_2025}). These features are prominent in cluster cores, indicating turbulent MFs and high thermal electron densities, while sources in the periphery (or projected onto it) typically show smoother RM distributions and higher degrees of polarisation (\citealt{Osinga_2022}, \citeyear{Osinga_2025}), consistent with weaker fields. 

CARLAJ1510 shows that the study of MF in proto-clusters is likely more complex than for local clusters. Using the polarised radiation from the central AGN, we found that its proto-ICM already hosts partially ordered MFs, including a coherent LOS MF component across most of the projected western lobe ($\sim 150$ kpc). This likely traces an external layer of the lobe or thin, compressed, weakly magnetised gas. On the other hand, the depolarisation of the eastern lobe suggests RM fluctuations on scales below our beam, likely within denser and more turbulent gas, as seen in evolved systems. The inferred lower limit of $0.4 \ \mu$G in the physical MF agrees with the predictions by cosmological MHD simulations, where weak seed fields (primordial or astrophysical) are amplified up to microgauss intensities by shocks and turbulence driven by accretion, mergers and feedback during large‑scale structure formation (e.g. \citealt{Ryu_2008}; \citealt{Vazza_2017}; \citealt{Donnert_2018}; \citealt{DF_2019}; \citealt{Tevlin_2025}). In addition, compressive motions by AGN jets may further enhance field strength and coherence, consistent with the QU fitting results presented in Sect. \ref{QU fitting results}. These findings are in agreement with the proto-cluster being dynamically young and still assembling, showing a mixture of turbulent and coherent MFs, possible asymmetries in ionised gas distribution, and ongoing interactions with the central radio galaxy. Such behavior aligns with simulations indicating that the early assembly of the hot gas phase at $z \sim 2-4$ is dominated by shock heating, gravitational collapse, and mergers \citep[e.g.][]{Barnes_2017, Remus_2023}. Supporting this scenario, \citet{Lepore_2024} combined X-ray and SZ observations of the Spiderweb proto-cluster. They identified a well-developed central cool core, likely fueling both intense star formation and the growth of the central supermassive black hole. A mild asymmetry in the X-ray emission was also observed, potentially indicating a cavity produced by radio jets or reflecting the dynamically evolving state of the system.
\vspace{-0.2cm}
\subsection{Depolarisation mechanisms} \label{Other scenarios}
A range of mechanisms can produce depolarisation, whether internal to the source or external along the LOS. These effects depend on a combination of factors, such as intrinsic source properties, morphology, environment, physical size, and viewing angle. Here, we discuss different scenarios and contributions to the observed polarisation and depolarisation.

As discussed in Sect. \ref{Medium Properties}, the strong polarisation asymmetry that we observe between the two AGN lobes (see Sect. \ref{RMS results}) provides insight into the surrounding medium. Such asymmetries have been reported for several decades in both low- and high-redshift radio galaxies. For instance, \citet{Pentericci_2000} used the JVLA at 4.7 and 8.2 GHz to measure the Faraday rotation in radio galaxies at $1.7 \lesssim z \lesssim 4.1$, and found RM differences between the two lobes (often several hundreds of $\rm rad\ m^{-2}$ in the source rest frame), with the more depolarised, higher RM lobe typically on the receding side of the jet. These features, along with disturbed morphologies, were interpreted as evidence for interactions with dense, magnetised environments such as proto-clusters. Similarly, a systematic JVLA study at 4.8 GHz of sources at $0.4 \lesssim z \lesssim 1$ by \citet{Goodlet_2004} reported consistent results, showing that depolarisation increases with redshift as the ambient medium of radio galaxies becomes denser and more complex. More recent Low Frequency Array (LOFAR)  observations at 150 MHz also reveal consistent asymmetries, but much weaker polarised emission compared to 1.4 GHz JVLA data, due to stronger external depolarisation \citep{Mahatma_2021, Piras_2025}. These trends are also observed in giant radio galaxies (GRGs), though their large sizes make them less affected by local environmental depolarisation, and therefore they are considered valuable probes of MFs and turbulence in the low-density intergalactic medium \citep{O'Sullivan_2018, O'Sullivan_2019, Stuardi_2020}.
\vspace{-0.2cm}
\subsubsection{Galactic halo contribution}
Since the AGN is inclined relative to the LOS, it is important to assess whether the host galaxy’s circumgalactic medium (CGM) contributes to the observed Faraday depolarisation. Although the host properties are unknown, its redshift and radio power suggest it may be a progenitor of a present-day brightest cluster galaxy. Such massive galaxies grow mainly through dry mergers and accretion \citep[e.g.][]{Laporte_2013, Zhao_2015}, with most of their mass already in place by $z>1$ \citep[e.g.][]{Rennehan_2020, Chu_2022}. The CGM is a multiphase medium, with a hot component ($10^6 \ \rm K$) that can extend to the virial radius and typically has higher X-ray luminosity in quiescent systems \citep[e.g.][]{FG_2023, Zhang_2024a, Zhang_2024b, Zhang_2025}. Its thermal content evolves with redshift, with cool gas dominating at high $z$ and hot haloes becoming prominent at lower $z$ \citep[e.g.][]{Huscher_2020, Wu_2025}. Both observations and simulations indicate that the CGM is magnetised (\citealt{Pakmor_2020}, \citeyear{Pakmor_2024}; \citealt{Lan_2020b}; \citealt{Bockmann_2023}). LOFAR studies show that the CGM of galaxies hosts weak, partially coherent MFs of a few $0.1\ \mu$G. In massive galaxies ($\rm M_*>10^{11}M_\odot$), the fields are small within $50-100$ kpc but increase mildly at larger distances, around $\sim0.8 r_{\rm vir}$ \citep{Carretti_2025b}, while in highly inclined galaxies coherent fields of similar strength are detected within $\sim100$ kpc \citep{Heesen_2023}. Further constraints come from radio galaxies using the polarisation asymmetry in the jets: \citet{Shah_2021} inferred small-scale fields of $\sim0.1-2.75\ \mu$G and much weaker ordered fields on $\sim100$ kpc scales for 57 elliptical galaxies up to $z \sim 0.5$; \citet{Seta_2021} estimated central RM dispersions of $\sim4.5-30$ rad m$^{-2}$, corresponding to central MF strengths of $0.1-1.3\ \mu$G (uniform density) up to $1.4-6.2\ \mu$G (King profile) for an historical sample of radio galaxies ($0.06 \lesssim z \lesssim 2.29$).\\
\\
Despite the difficulty of separating CGM and proto-ICM contributions at high redshift, we estimated the galactic halo contribution following \citet{Seta_2021}. Assuming a King density profile, $\rm n_e(r) = n_0 \left[ 1 + (r/a)^2 \right]^{-3\beta/2}$, with $\rm B(r) \propto n_e(r)^\gamma$ and adopting $\beta = 1/2$ and $\gamma = 2/3$, the RM dispersion as a function of projected distance $\rm r_\perp$ becomes $\rm \sigma_{\rm RM}(r_\perp) = \sigma_{\rm RM}(0) \left[ 1 + (r_\perp/a)^2 \right]^{-0.22}$. Using a core radius $a = 3$ kpc and a central dispersion $\sigma_{\rm RM}(0) = 20.3 \pm 0.8$ rad m$^{-2}$ (for $1.5 < z < 2.0$ galaxies; \citep{Seta_2021}), we found $\sigma_{\rm RM} = 4.8 \pm 0.2$ rad m$^{-2}$ at a projected distance of 80 kpc, roughly the projected extent of the eastern lobe. These estimates suggest that the host CGM provides only a minor enhancement to Faraday dispersion, with depolarisation dominated by the dense, turbulent proto-ICM.
\vspace{-0.3cm}
\subsubsection{Internal depolarisation}
While the polarisation asymmetry in the radio lobes can be explained by an external Faraday screen, some aspects of the data suggest that internal depolarisation may also contribute. Internal depolarisation arises when synchrotron-emitting and Faraday-rotating plasma coexist, so that different regions rotate the polarisation angle by different amounts, reducing the net observed polarisation. The frequency-dependent depolarisation then deviates from a simple exponential law and depends on the MF structure, whether turbulent or ordered \citep[e.g.][]{Burn_1966, Sokoloff_1998, Yushkov_2024}.

In the literature, internal Faraday depolarisation has been reported in several radio sources. For instance, \citet{Pasetto_2018} found that most of the AGN in their JVLA $1-12$ GHz sample require multiple Faraday components, often associated with local turbulent plasma or mixed with the emitting region, while only two are consistent with simple differential rotation in an ordered MF. In spatially resolved jets, internal depolarisation can be constrained even more robustly: \citet{Pasetto_2021} used $4-8$ GHz JVLA imaging to reveal a double-helix morphology with transverse RM gradients in M87, consistent with a helical MF. High-resolution VLBA studies have likewise found jets requiring internal Faraday rotation \citep[e.g.][]{Hovatta_2012, Molina_2014, Kravchenko_2017}, although external screens usually dominate in larger samples. Furthermore, asymmetries in the polarisation properties of the two radio jets may arise from intrinsic physical properties, including MF topology, relativistic effects, and plasma composition (e.g. \citealt{Lyutikov_2005}; \citealt{Calusen_Brown_2011}).}

In our case, the low RM dispersion ($36\ \mathrm{rad\ m^{-2}}$) and coherent projected MF vectors suggest that the advancing western lobe may either order the MF in its outer regions or compress a magnetised external medium. In this scenario, differential Faraday depolarisation arising from an ordered field can occur without fully depolarising the lobe. This is supported by the Faraday complexity obtained from the QU fitting results (model n3; see Sect. \ref{QU fitting results}), which are consistent with synchrotron emission and Faraday rotation arising in the same plasma, with additional depolarisation from the proto-ICM ($\sigma_{\rm RM,ext}= 63.6 \pm 0.7 \ \rm rad \ m^{-2}$). However, in our case, the jet is unresolved, and the $7^{\prime\prime}$ beam averages emission from multiple regions, preventing us from resolving spatially the thick Faraday component. Therefore, higher-frequency observations are needed to investigate such internal Faraday rotation and to model it accurately. For instance, JVLA observations in A configuration would provide the angular resolution required to probe the source structure on smaller scales, paving the way for future studies with the ngVLA \citep{DiFrancesco_2019}.
\vspace{-0.2cm}
\subsection{Limitations on the observations and MF modelling} \label{Limitations}
Several factors affect the interpretation of our polarimetric results and the constraints derived on the MF in the proto-ICM. The first major source of uncertainty in our MF estimates comes from the gas density and its distribution along the LOS. Determining the thermal electron density is crucial to interpret the observed RM and infer the MF properties. In principle, X-ray or SZ observations provide this information, but at $z > 1.5$ the low brightness and AGN contamination make such measurements challenging. So far, the Spiderweb complex is the only proto-cluster at comparable redshift with reliable X-ray \citep{Tozzi_2022b} and SZ \citep{Di_Mascolo_2023} detections of diffuse ICM emission within 100 kpc. Alternatively, cosmological simulations can provide useful priors. In our analysis, we assumed a simplified model for the proto-ICM and AGN jet orientations (Fig. \ref{fig:Sketch}), but variations in inclination or gas distribution can significantly affect the RM pattern; for instance, the lack of polarisation in the eastern lobe may result from these asymmetries or local density enhancements along the LOS.

The JVLA L band provides good Faraday resolution ($60\ \mathrm{rad \ m^{-2}}$), allowing the separation of multiple Faraday components along the LOS. This requires the inclusion of the low-frequency channels, which set the synthesised beam to $7^{\prime\prime}\ \times7^{\prime\prime}$ ($\sim 59$ kpc). At this resolution, depolarisation effects become stronger, especially in turbulent magnetised media with fluctuations on scales smaller than the beam. Because of the vector nature of polarisation (Eq. \ref{eq:P}), small-scale MF fluctuations within the beam reduce the observed signal (beam depolarisation; \citealt{Burn_1966}). This behaviour is evident in our data, where the polarised intensity decreases in regions of enhanced turbulence. Thus, while low-frequency coverage is essential for probing the Faraday structure, it limits the recovery of the intrinsic polarisation, and high-frequency observations are required to mitigate depolarisation. Nonetheless, the current frequency range remains well suited for detecting the RM values expected in the medium of high-redshift systems (e.g. \citealt{Rappaz_2024}).

For the 3D simulations, we adopted simplified models for the gas density and MF power spectrum, based on low-redshift clusters, where the MF strength scales with the gas density \citep[e.g.][]{Bonafede_2010}. The gas was modelled as a uniform density cubic volume, with values from preliminary cosmological simulations. Proto-cluster environments are, however, more dynamically evolving and may deviate from these assumptions. Since, the MF power spectrum at high redshift is unknown, we assumed a Kolmogorov spectrum, as commonly used for low-$z$ systems. Future simulations using resolved density profiles, combined with multiwavelength observations, will be required to better characterise MFs in these early structures.

Additional constraints could come from background polarised sources probing multiple LOS across the proto-ICM, tracing RM and LOS MF variations. Looking ahead, the SKA Observatory (SKAO) will offer an unprecedented opportunity in this field. The combination of SKAO-Mid Band 1 ($350 - 1050$ MHz) and Band 2 ($950-1760$ MHz) will provide high sensitivity, Faraday resolution, and Faraday depth coverage, allowing the detection of numerous faint polarised sources and precise RM measurement, opening a new window in our understanding of cosmic magnetism.

\section{Summary and conclusions} \label{Conclusions}
In this work we presented a polarimetric study of the high-redshift proto-cluster CARLAJ1510 at $z = 1.72$. We used JVLA L band ($1-2$ GHz) observations to investigate the MF properties (intensity and configuration) of the proto-ICM during the early stages of cluster assembly.   

The data were processed in both continuum and polarisation, updating the calibrators models provided by \citealt{Taylor_2024} for the total intensity, and by \citealt{Hugo_2024} for the polarised intensity and polarisation angle. To investigate the MF in the proto-ICM, we focused the analysis on a region of $\sim 1$ Mpc, encompassing all six spectroscopically confirmed members, and analysed the Faraday rotation effect on the polarised emission from the central AGN jets and lobes. We applied the RM synthesis technique and produced masked maps (thresholded at $6\sigma_{\mathrm{P,i}}$) of the maximum polarised intensity, RM, and fractional polarisation. In the western lobe, we found a uniform RM distribution dominated by negative values ($\rm \langle RM\rangle = -115 \pm 32 \ rad \ m^{-2}$ in the source frame), indicating a LOS MF component directed away from the observer. The low RM dispersion ($\rm \sigma_{RM} = 36 \pm 11 \ rad \ m^{-2}$ in the source frame) and the coherent projected MF vectors suggest a partially ordered MF across most of the projected western lobe
 ($\sim 150$ kpc). This region also exhibits a significant drop in the fractional polarisation (from 8.2\% to 0.3\%), while the eastern lobe is almost completely depolarised. We interpret these features as being due to an external region of the lobe or to the AGN compressing a nearby thin magnetised layer, together with being embedded in a turbulent, magnetised plasma responsible for the depolarisation. 
 
To further investigate these features, we performed a QU fitting analysis of the polarised data. This suggests the presence of an internal Faraday-thick component, likely associated with the lobe’s magnetised plasma, that is mixed with the surrounding thermal gas and is possibly actively magnetising the proto-ICM.
 
To interpret the observations, we performed 3D simulations modelling the gas density and a random MF with different strengths and auto-correlation lengths. By assuming different AGN orientations, we generated mock RM maps and compared them with the observations. The results show that the Faraday rotation in the western lobe requires an ordered MF component and cannot be reproduced by a purely turbulent field. Consequently, the RM values from this region cannot directly constrain the field strength using this model. However, assuming that the two lobes are intrinsically similar, the non-detection of the eastern lobe's polarisation allowed us to set lower limits on the physical MF intensity. Depending on the assumed MF power spectrum, we derived values ranging from a few microgauss, and decreasing with increasing path length. From this, we place a conservative lower limit of $0.4 \ \mu$G, consistent with the upper limit reported by \citet{Anderson_2022} for the Spiderweb proto-cluster.

We note that the lower limit we have derived is based on several assumptions. While the scenario in which depolarisation originates from external sources is robust, the gas density, its distribution, and the geometry of the MF depend on the specific model used. Nonetheless, our analysis proves for the first time the presence of non-zero MF in the early phases of cluster formation, suggesting that the magnetisation of the ICM begins at high redshift, with partially ordered fields shaped by AGN feedback. 

Our results are consistent with predictions from cosmological simulations and reveal a more complex picture compared to low-redshift evolved clusters, reflecting the dynamically young state of CARLAJ1510.  Overall, our study provides the first direct constraints on MFs in a high-redshift proto-cluster environment, and  bridges the gap between primordial seed fields and the well-established microgauss fields in local clusters.

\begin{acknowledgements}
    This paper makes use of data from the Karl G. Jansky Very Large Array. The National Radio Astronomy Observatory is a facility of the National Science Foundation operated under cooperative agreement by Associated Universities, Inc. AP, AB and DVP acknowledge support from the ERC CoG \vec{B}ELOVED, GA n. 101169773. CJR acknowledges support from the DFG via the Collaborative Research Center SFB1491 \textit{Cosmic Interacting Matters - From Source to Signal} (project no.\ 445052434). CS acknowledges support from  the Fondazione ICSC, Spoke 3 Astrophysics and Cosmos Observations - National Recovery and Resilience Plan (Piano Nazionale di Ripresa e Resilienza, PNRR) Project ID CN00000013 “Italian Research Center for High-Performance Computing, Big Data and Quantum Computing” funded by MUR Missione 4 Componente 2 Investimento 1.4: Potenziamento strutture di ricerca e creazione di “campioni nazionali di R\&S (M4C2-19)” – Next Generation EU (NGEU).
    \vspace{-0.3cm}
\end{acknowledgements}
\bibliography{mybibliography}
\bibliographystyle{aa}
\begin{appendix} 
\onecolumn
\section{3C286 total intensity and polarisation models} \label{3C286 model}
\begin{figure}[!htb]
    \vspace{-0.3cm}
    \begin{subfigure}
        \centering
        \includegraphics[width=0.33\linewidth]{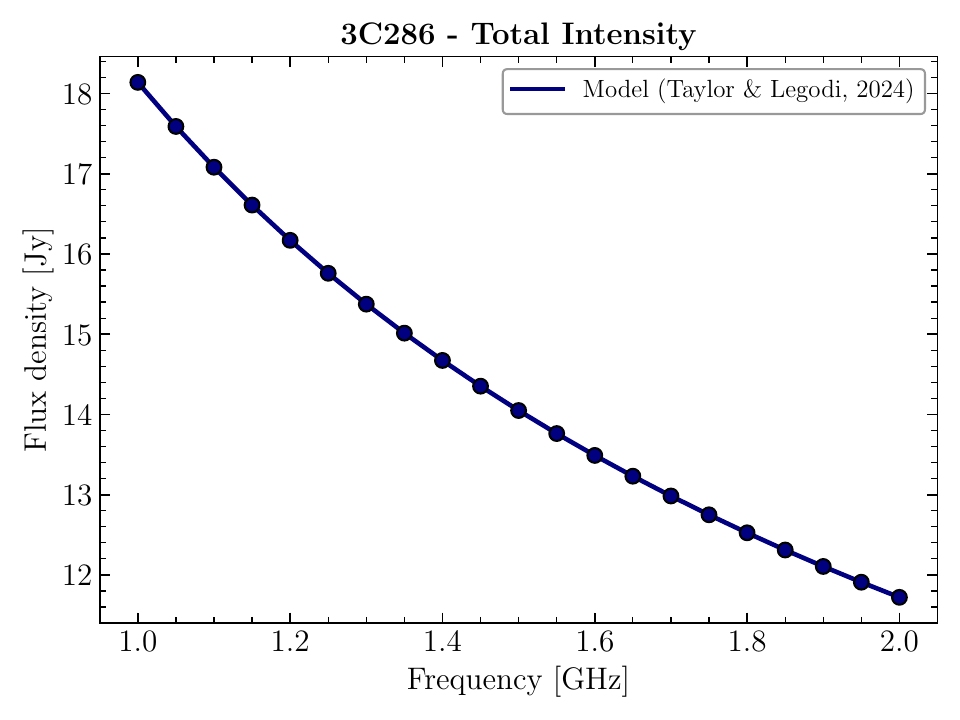}
    \end{subfigure}
    \begin{subfigure}
        \centering
        \includegraphics[width=0.33\linewidth]{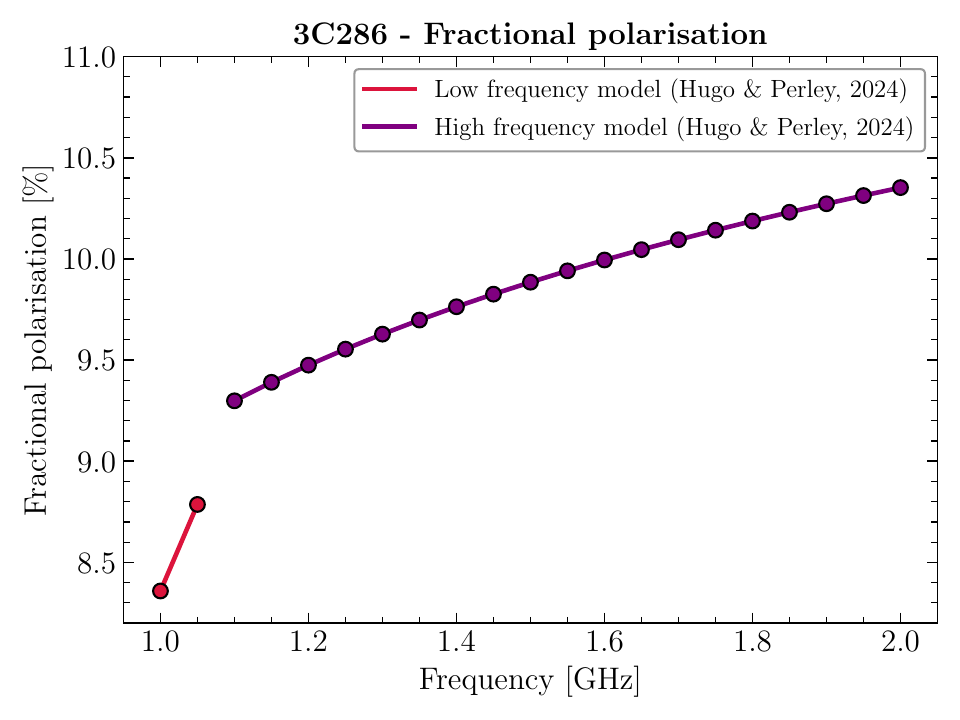}
    \end{subfigure}
    \begin{subfigure}
        \centering
        \includegraphics[width=0.33\linewidth]{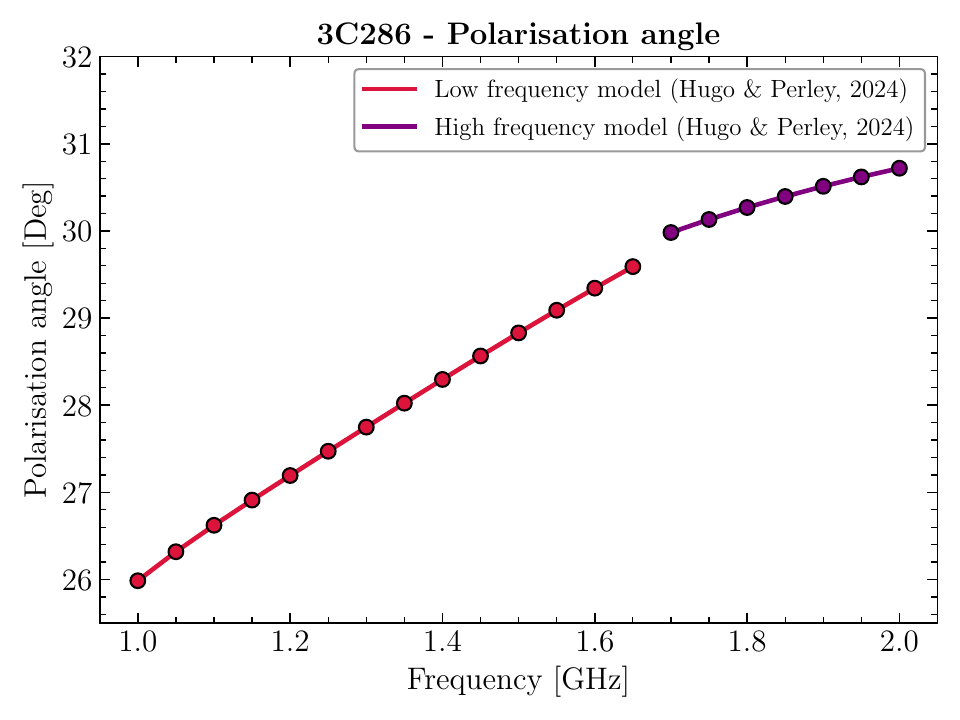}
    \end{subfigure}
    \vspace{-0.5cm}
    \caption{Models for the primary calibrator 3C286 between $1-2$ GHz. Left: Total intensity. Centre: Fractional polarisation. Right: Polarisation angle. The total intensity spectrum is shown according to \citealt{Taylor_2024} model. The red and purple curves in the polarisation plots represent the low- and high-frequency models of \citealt{Hugo_2024}, respectively. The markers indicate the model-sampled values.}
    \label{fig:3C286models}
\end{figure}

In this section, we present the model of the primary calibrator 3C286 in total intensity, fractional polarisation, and polarisation angle, following \citealt{Taylor_2024} and \citealt{Hugo_2024}. It is important to note that the original models were derived from MeerKAT observations. While the MeerKAT L band differs slightly from the JVLA L band, these models have been adapted and applied for calibration (Sect. \ref{Calibration}) in the JVLA 1-2 GHz frequency range. Plots of the models are shown in Fig. \ref{fig:3C286models}.\\
\\
For the total intensity, we adopted the curved power-law model presented by \citet{Taylor_2024} and expressed as:
\begin{center}
$\rm I(\nu) = I_0 \left( \frac{\nu}{\nu_0} \right)^{\alpha + C \ln(\nu/\nu_0)}$,
\end{center}
where $\rm I_0$ is the intensity at the reference frequency, $\alpha$ is the spectral index, and C is the curvature term.\\
\\
For the fractional polarisation and polarisation angle, we, instead, used the broken power-law models presented by \citealt{Hugo_2024}, expressed respectively as:
\begin{center}
    $\rm P(\nu_\mathrm{GHz}) =
\begin{cases} 
0.080 - 0.053 \, \lambda^2 - 0.015 \log_{10} \lambda^2, & \nu_\mathrm{GHz} \in [1.1, 12] \\[1mm]
0.029 - 0.172 \, \lambda^2 - 0.067 \log_{10} \lambda^2, & \nu_\mathrm{GHz} < 1.1,
\end{cases}$
\end{center}
\begin{center}
    $\mathrm{EVPA}(\nu_\mathrm{GHz}) [\mathrm{deg}] =
\begin{cases} 
32.64 - 85.37 \, \lambda^2, & \nu_\mathrm{GHz} \in [1.7, 12] \\[2mm]
29.53 + \lambda^2 \left( 4005.88 \log_{10}^3 \nu_\mathrm{GHz} - 39.38 \right), & \nu_\mathrm{GHz} < 1.7.
\end{cases}$
\end{center}
Here $\lambda$ is the wavelength in m and $\nu$ is the frequency in GHz.\\
\\
To fit both the frequency-dependent trends of the fractional polarisation (P) and the linear polarisation angle (EVPA), we used sixth-order polynomials around 1.5 GHz:
\begin{center}
    $\mathrm{P}(\nu_{\mathrm{GHz}}) = a_0 + a_1 x + a_2 x^{2} + \dots + a_6 x^{6}$, \ \ \ $\mathrm{PA}(\nu_{\mathrm{GHz}}) = b_0 + b_1 x + b_2 x^2 + \dots + b_6 x^6$, \ \ \  $x = \frac{\nu_{\mathrm{GHz}} - 1.5}{1.5}$.
\end{center}
The coefficients $a_i$ and $b_i$ ($i=0,\dots,6$) were determined by fitting the polynomials to the model values at the observed frequencies using the function \texttt{curve\_fit} from the \texttt{scipy.optimize} module in Python.\\
\\
Using the \texttt{setjy} task, we set up the models by specifying the parameters and polynomial coefficients given in Table \ref{tab:Calparameters}. 
\begin{table}[!htb]
    \centering
    \caption{Reference model parameters for 3C286 and polynomial coefficients for fractional polarisation and polarisation angle.}
    \label{tab:3C286_model}
    \renewcommand{\arraystretch}{1.8}
    \begin{tabular}{l l }
        \hline
        \noalign{\smallskip}
        Stokes I model parameters & \hspace{0.3cm}
    \begin{tabular}{c c c c c c c c c c c}
         & &  I & Q & U & V & $\nu_\mathrm{ref}$ & $\alpha$ & $C$ & & \\
         & &  {[Jy]} & {[Jy]} & {[Jy]} & {[Jy]} & {[GHz]} & & & & \\
        \hline
         & & 14.05 & 0 & 0 & 0 & 1.5 & -0.63 & 0.001 & & 
    \end{tabular} \\
    \hline
    Fractional polarisation $a_i$ (i = 0,$\dots$,6) [\%] & 
    \begin{tabular}{c c c c c c c}
        0.0988 & 0.0192 & -0.0132 & -0.1409 & 0.2891 & 2.1591 & -5.4740
    \end{tabular} \\
    \hline
    Polarisation angle $b_i$ (i = 0,$\dots$,6) [rad] & 
    \begin{tabular}{c c c c c c c}
        0.5031 & 0.1463 & 0.0309 & -0.3612 & -2.0157 & 1.4189 & 9.7180
    \end{tabular} \\
    \hline
    \label{tab:Calparameters}
    \end{tabular}
    \vspace{-1cm}
\end{table}
\section{QU fitting models and initial QU fitting analysis} \label{QU fitting-2}
\subsection{QU fitting models} \label{QU fitting model}
This section summarises the polarisation models tested in the QU fitting analysis (Sect. \ref{QU fitting results} and Sect. \ref{QU fit results}), where the complex fractional polarisation (Eq. \ref{eq:P}) is modelled as a function of $\lambda^2$. The models account for both single and multiple Faraday components, differential and external Faraday rotation, including both internal and external depolarisation affecting the emitting region. 
\subsubsection{Model parameters}
\begin{itemize}
    \itemsep3pt
    \item $\lambda$: Observing wavelength
    \item $\rm \psi_0$, $\rm \psi_{0,i}$: Intrinsic polarisation angle(s)
    \item $\rm p_i$, $\rm p_{i,j}$: Intrinsic fractional polarisation(s)
    \item RM, $\rm RM_i$: Faraday RM(s) in $\rm rad \ m^{-2}$
    \item $\sigma_{\rm RM}$: Faraday dispersion in $\rm rad \ m^{-2}$
    \item $\rm \Delta RM$, $\rm \Delta RM_i$: Differential Faraday rotation across a slab in $\rm rad \ m^{-2}$
    \item $\rm RM_{\rm screen}$, $\sigma_{\rm RM, screen}$: Rotation measure and dispersion due to an external screen in $\rm rad \ m^{-2}$
    \item S: Complex depolarisation term defined in model m7 and m12
\end{itemize}
\subsubsection{Polarisation models}
\begin{itemize}
    \item m1: Single Faraday-simple source
\end{itemize}
\begin{center}
$\rm p = p_i\ \cdot \  \exp[2i \ (\psi_0+RM\lambda^2)] \label{eq:m1}$
\end{center}
\begin{itemize}
    \item m2: Single Faraday component with external Faraday dispersion
\end{itemize}
\begin{center}
$\rm p = p_i\ \cdot \  \exp[2i \ (\psi_0+RM\lambda^2)] \ \cdot \ \exp[-2\sigma_{RM}^2\lambda^4] \label{eq:m2}$
\end{center}
\begin{itemize}
    \item m3: Two Faraday components with a common external Faraday dispersion term
\end{itemize}
\begin{center}
$\rm p =\{p_{i,1}\ \cdot \  \exp[2i \ (\psi_{0,1}+RM_1\lambda^2)] + p_{i,2}\ \cdot \ \exp[2i \ (\psi_{0,2}+RM_2\lambda^2)]\} \ \cdot \ \exp[-2\sigma_{RM}^2\lambda^4] \label{eq:m3}$
\end{center}
\begin{itemize}
    \item m5: Single Faraday component with differential Faraday rotation (Burn slab)
\end{itemize}
\begin{center}
$\rm p =  p_i \cdot \exp\left[2i \left( \psi_0 + \left(RM + \frac{1}{2} \Delta RM \right) \lambda^2 \right)\right] \cdot \frac{\sin(\Delta RM \cdot \lambda^2)}{\Delta RM \cdot \lambda^2} \label{eq:m5}$
\end{center}
\begin{itemize}
    \item m6: Two Faraday components with differential Faraday rotation (double Burn slab)
\end{itemize}
\begin{center}
$\rm p= \Bigl\{p_{i,1} \cdot \exp\left[2i \left( \psi_{0,1} + \left(RM_1 + \frac{1}{2} \Delta RM_1 \right) \lambda^2 \right)\right] \cdot  \frac{\sin(\Delta RM_1 \cdot \lambda^2)}{\Delta RM_1 \cdot \lambda^2}\Bigr\} + \Bigl\{p_{i,2} \cdot \exp\left[2i \left( \psi_{0,2} + \left(RM_2 + \frac{1}{2} \Delta RM_2 \right) \lambda^2 \right)\right] \cdot 
\frac{\sin(\Delta RM_2 \cdot \lambda^2)}{\Delta RM_2 \cdot \lambda^2}\Bigr\} \label{eq:m6}$
\end{center}
\begin{itemize}
    \item m7: Single Faraday component with internal Faraday dispersion
\end{itemize}
\begin{center}
$\rm p = p_i\ \cdot \ \exp[2i \ (\psi_0+RM\lambda^2)] \ \cdot \ \left\{\frac{1-\exp(-S)}{S}\right\} \label{eq:m7}$,\\
\vspace{0.2cm}
\justifying
where $\rm S= 2\lambda^4\sigma_{RM}^2 - 2i\lambda^2\Delta_{RM}$ and $\rm \Delta_{RM}$ is the total Faraday depth through the region.
\end{center}

\begin{itemize}
    \item m12: Single Faraday component with internal Faraday dispersion and a foreground external dispersion screen
\end{itemize}
\begin{center}
$\rm p = p_i\ \cdot \ \exp[2i \ (\psi_0+RM_{\rm screen}\lambda^2)] \ \cdot \ \exp[-2\sigma_{RM, screen}^2\lambda^4] \ \cdot \ \left\{\frac{1-\exp(-S)}{S}\right\} \label{eq:m12}$
\end{center}
\subsubsection{New polarisation models}
\begin{itemize}
    \item n1: Three Faraday components with a common external Faraday dispersion term
\end{itemize}
\begin{center}
$\rm p =\{p_{i,1}\ \cdot \  \exp[2i \ (\psi_{0,1}+RM_1\lambda^2)] + p_{i,2}\ \cdot \ \exp[2i \ (\psi_{0,2}+RM_2\lambda^2)] + p_{i,3}\ \cdot \ \exp[2i \ (\psi_{0,3}+RM_3\lambda^2)]\} \ \cdot \ \exp[-2\sigma_{RM}^2\lambda^4] \label{eq:n1}$
\end{center}
\begin{itemize}
    \item n2: Three Faraday components with differential Faraday rotation (triple Burn slab)
\end{itemize}
\begin{center}
$\rm p= \Bigl\{p_{i,1} \cdot \exp\left[2i \left( \psi_{0,1} + \left(RM_1 + \frac{1}{2} \Delta RM_1 \right) \lambda^2 \right)\right] \cdot  \frac{\sin(\Delta RM_1 \cdot \lambda^2)}{\Delta RM_1 \cdot \lambda^2}\Bigr\} + \Bigl\{p_{i,2} \cdot \exp\left[2i \left( \psi_{0,2} + \left(RM_2 + \frac{1}{2} \Delta RM_2 \right) \lambda^2 \right)\right] \cdot 
\frac{\sin(\Delta RM_2 \cdot \lambda^2)}{\Delta RM_2 \cdot \lambda^2}\Bigr\} + \Bigl\{\rm p_{i,3} \cdot \exp\left[2i \left( \psi_{0,3} + \left(RM_3 + \frac{1}{2} \Delta RM_3 \right) \lambda^2 \right)\right] \cdot 
\frac{\sin(\Delta RM_3 \cdot \lambda^2)}{\Delta RM_3 \cdot \lambda^2}\Bigr\} \label{eq:n2}$
\end{center}
\begin{itemize}
    \item n3: One Faraday-simple component, one Faraday component with differential Faraday rotation and a common external Faraday dispersion term
\end{itemize}
\begin{center}
$\rm p = \Bigl\{p_{i,1} \cdot \exp\left[2i \left( \psi_{0,1} + RM_1 \cdot \lambda^2 \right) \right]  + p_{i,2} \cdot  \exp\left[ 2i \left( \psi_{0,2} + \left(RM_2 + \frac{1}{2} \Delta RM_2 \right) \lambda^2 \right) \right] \cdot \frac{\sin(\Delta RM_2 \cdot \lambda^2)}{\Delta RM_2 \cdot \lambda^2}\Bigr\} \cdot \exp\left( -2 \sigma^2_{RM} \cdot \lambda^4 \right) \label{eq:n3}$
\end{center}
\subsection{Initial model fitting results} \label{QU fit results}
In this section, we present the results of the initial QU fitting attempts. Several polarisation models have been tested (see Appendix \ref{QU fitting model} for details), and examples of spectral fit and corner plots of some of them are shown in Appendix \ref{QUfitting plots}. In the initial analysis, we excluded the presence of multiple Faraday components and focused on single-component models to distinguish between internal and external depolarisation. Specifically, we considered: a model without depolarisation (m1), one with external depolarisation (m2), one including differential Faraday rotation (m5), a model with internal depolarisation (m7), and one combining internal and external depolarisation (m12). 
Some of these models can reproduce the average RM within the uncertainties and some others may also suggest the presence of external depolarisation, but the reduced chi-squared, $\chi^2_\nu$, is always larger than 24.  Moreover, some models are able to reproduce the fractional Stokes Q and U at low frequencies, although they fail to capture the complex high-frequency behaviour. This pattern may suggest the presence of multiple intervening Faraday components. We therefore explored models that account for them, for example model m3, which includes two Faraday-simple components with common external depolarisation, and m6, which includes two Faraday components undergoing differential Faraday rotation. To investigate further, we introduced three additional models, n1, n2 and n3: model n1 extends m3 by adding a third Faraday-simple component, while n2 includes three components with differential Faraday rotation; model n3, instead, includes two components, one Faraday-simple and one with differential Faraday rotation, both subject to external depolarisation. The latter is discussed in Sect. \ref{QU fitting results}. A summary of the statistics for all models is provided in Table \ref{tab:QUfitstats}.
\vspace{-0.2cm}
\begin{table}[!htb]
    \centering
    \caption{Statistics of the QU fitting for each model.}
    \label{tab:QUfitstats}
    \begin{tabular}{c c c c | c c c c}
        \hline
        \noalign{\smallskip}
        \noalign{\smallskip}
        Model &  & $\chi^2_{\nu}$ & & & Model &  & $\chi^2_{\nu}$ \\
        (1) & & (2) & & & (1) & & (2) \\
        \noalign{\smallskip}
        \noalign{\smallskip}
        \hline
        \noalign{\smallskip}
        \noalign{\smallskip}
        m1 & & 62.1 & & & m7 & & 25.6\\[4pt] 
        m2 & & 25.7 &  & & m12 & & 25.9\\[4pt] 
        m3 & & 18.9 & & & n1 & & 26.3 \\[4pt]
        m5 & & 24.9 & & & n2 & & 24.0\\[4pt]
        m6 & & 15.7 & & & n3 & & 14.7 \\
        \noalign{\smallskip}
        \hline
    \end{tabular}
    \tablefoot{(1): Name of the model; (2): Reduced chi-squared of the QU fitting.}
\end{table}
\vspace{-0.5cm}
\subsection{Spectral fit and corner plots of QU fitting models} \label{QUfitting plots}
\begin{figure}[!htb]
    \centering
    \begin{subfigure}
        \centering
        \includegraphics[width=0.525\linewidth]{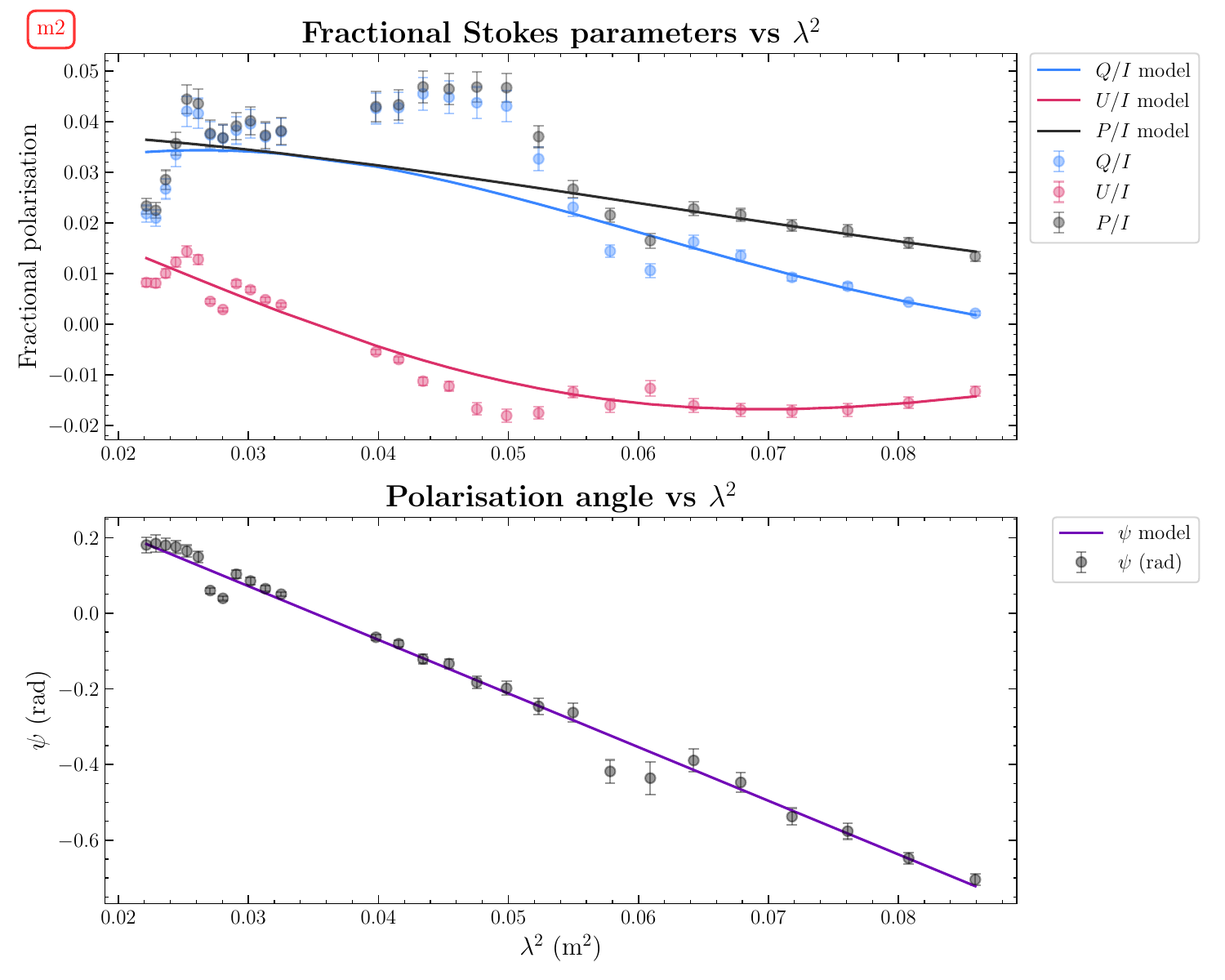}
    \end{subfigure}
    \begin{subfigure}
        \centering
        \includegraphics[width=0.425\linewidth]{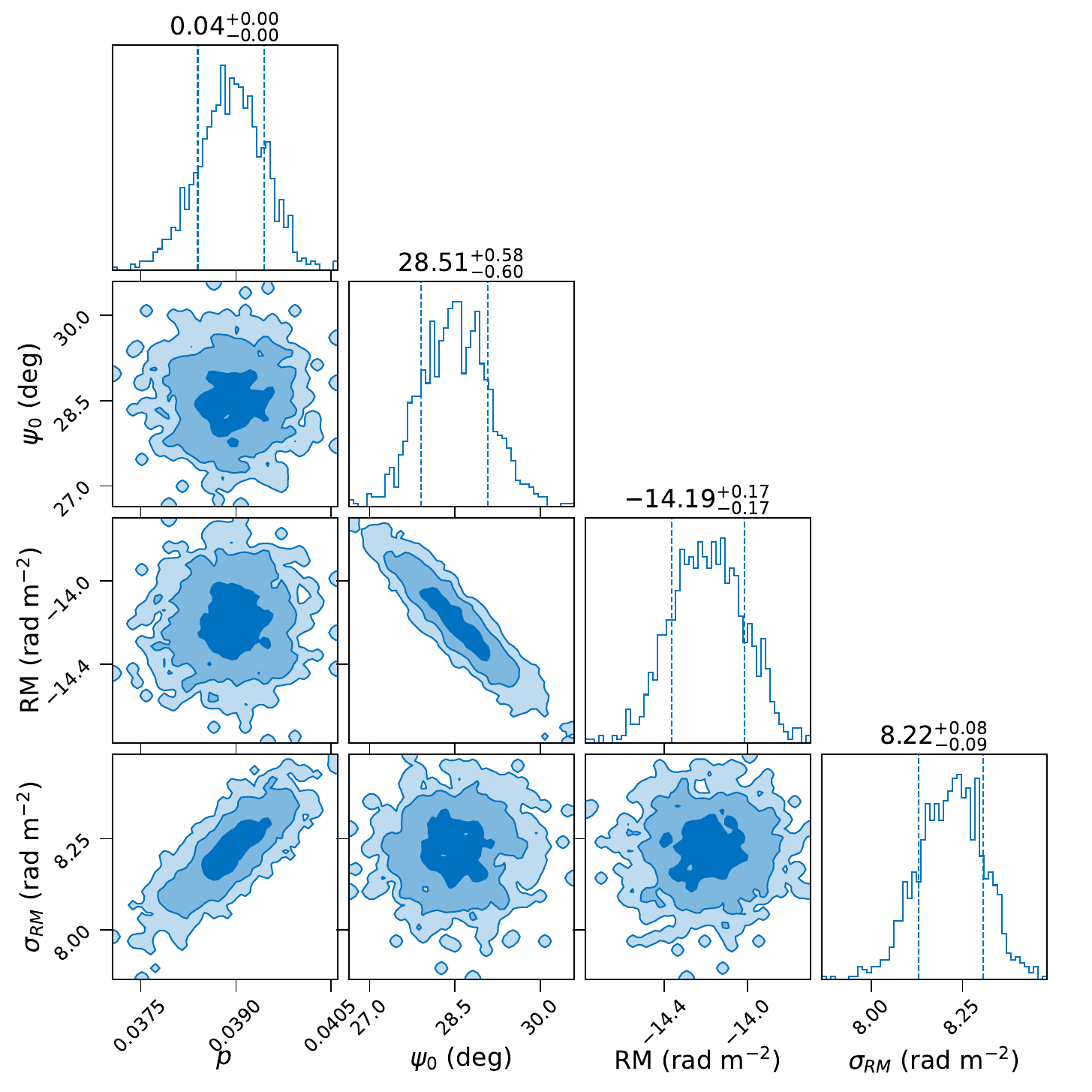}
    \end{subfigure}    
    \caption{QU fitting plots for model m2. Left: Best fit of the fractional polarisation (P/I, Q/I, and U/I) and polarisation angles as a function of $\lambda^2$. The data points represent the observations;  the straight lines are the model predictions (P/I: black, Q/I: blue, U/I: red, $\psi$: purple). Right: Probability distributions of all the model parameters: fractional polarisation ($p$), intrinsic polarisation angle ($\psi_{0}$ in deg), RM in $\rm rad \ m^{-2}$, and $\sigma_{\rm RM}$ in $\rm rad \ m^{-2}$.}
    \label{fig:QUfit-m2}
\end{figure}

\begin{figure}[!htb]
    \vspace{1.5cm}
    \centering
    \begin{subfigure}
        \centering
        \includegraphics[width=0.525\linewidth]{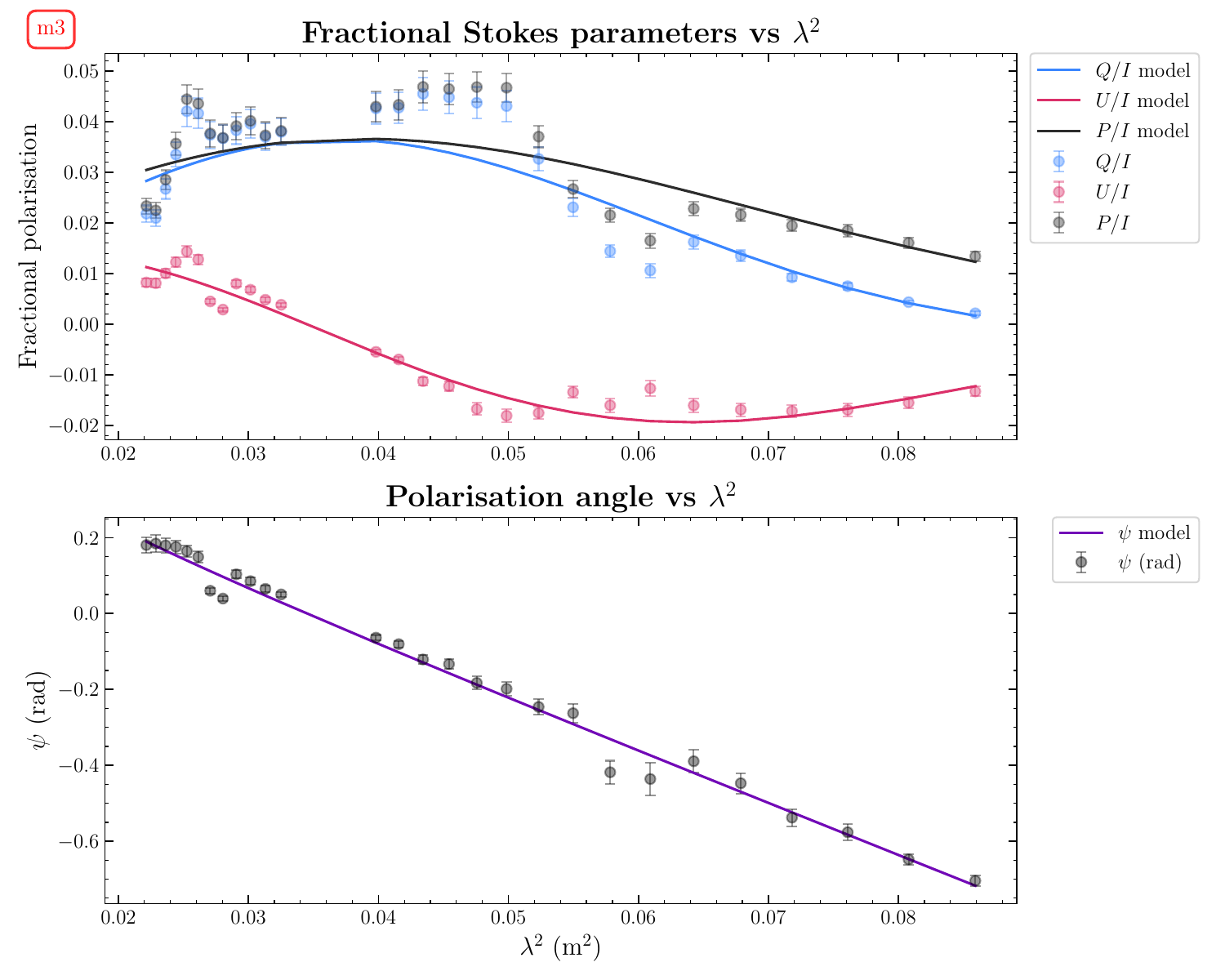}
    \end{subfigure}
    \begin{subfigure}
        \centering
        \includegraphics[width=0.45\linewidth]{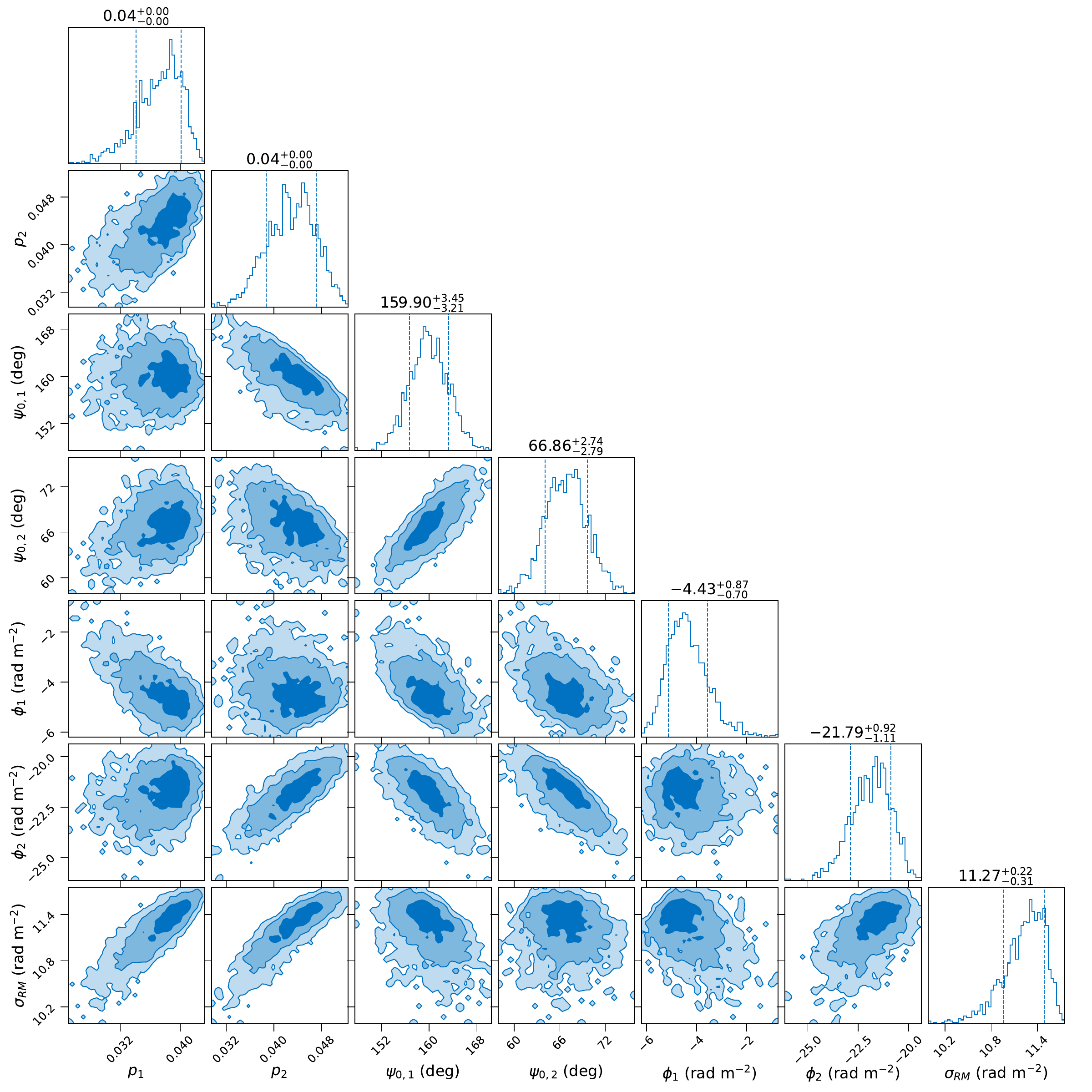}
    \end{subfigure}    
    \caption{QU fitting plots for model m3. Left: Best fit of the fractional polarisation (P/I, Q/I, and U/I) and polarisation angles as a function of $\lambda^2$. The data points represent the observations;  the straight lines are the model predictions (P/I: black, Q/I: blue, U/I: red, $\psi$: purple). Right: Probability distributions of all the model parameters: fractional polarisation ($p_1$ and $p_2$), intrinsic polarisation angle ($\psi_{0,1}$ and $\psi_{0,2}$ in deg), Faraday depth ($\phi_1$ and $\phi_2$ in $\rm rad \ m^{-2}$), and external Faraday dispersion ($\sigma_{\rm RM}$ in $\rm rad \ m^{-2}$).}
    \label{fig:QUfit-m3}
\end{figure}

\begin{figure}[!htb]
    \vspace{1cm}
    \centering
    \begin{subfigure}
        \centering
        \includegraphics[width=0.525\linewidth]{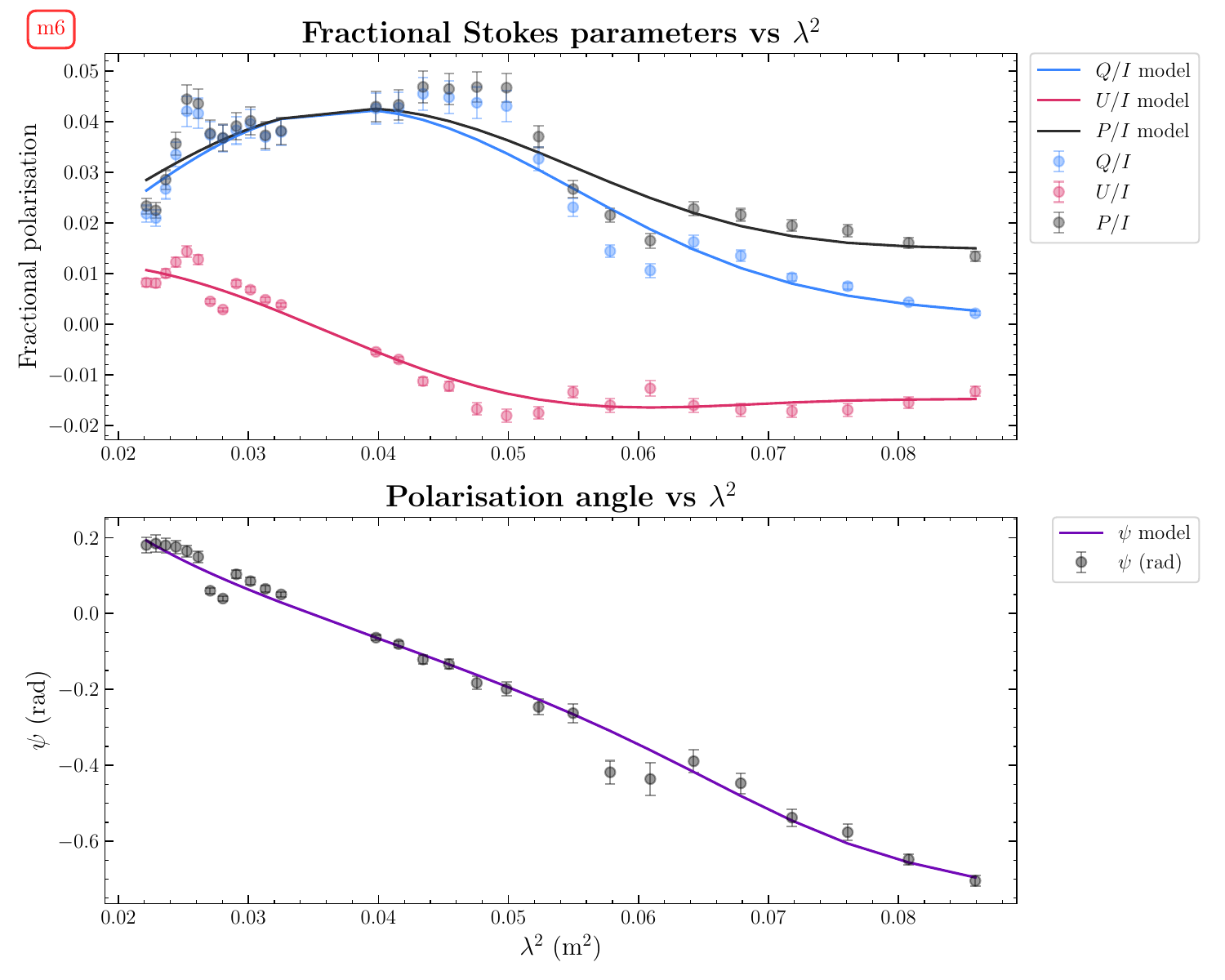}
    \end{subfigure}
    \begin{subfigure}
        \centering
        \includegraphics[width=0.445\linewidth]{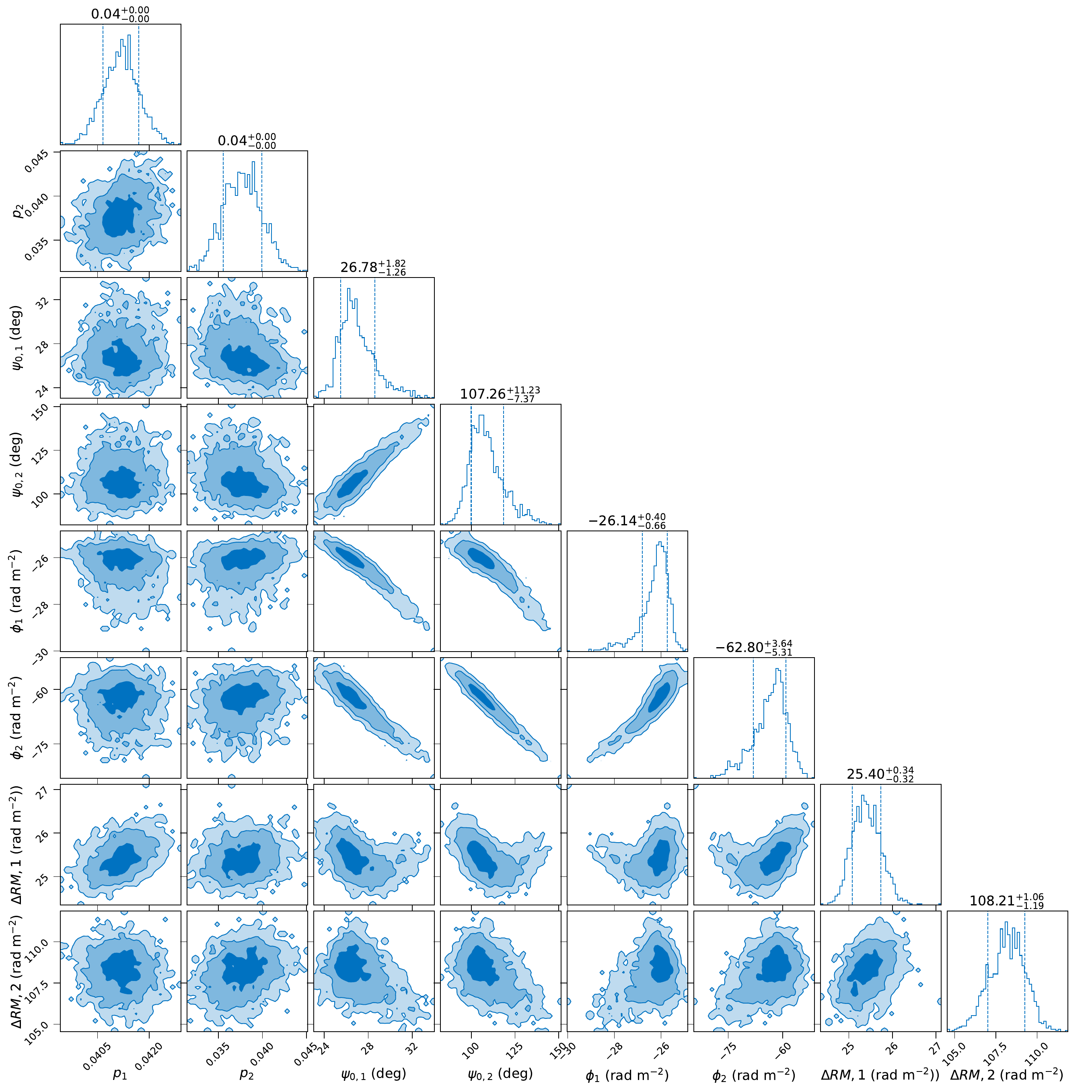}
    \end{subfigure}    
    \caption{QU fitting plots for model m6. Left: Best fit of the fractional polarisation (P/I, Q/I, and U/I) and polarisation angles as a function of $\lambda^2$. The data points represent the observations;  the straight lines are the model prediction (P/I: black, Q/I: blue, U/I: red, $\psi$: purple). Right: Probability distributions of all the model parameters: fractional polarisation ($p_1$ and $p_2$), intrinsic polarisation angle ($\psi_{0,1}$ and $\psi_{0,2}$ in deg), mean RM ($\phi_1$ and $\phi_2$ in $\rm rad \ m^{-2}$), and internal Faraday dispersion ($\Delta \rm RM_1$ and $\Delta \rm RM_2$ in $\rm rad \ m^{-2}$).}
    \label{fig:QUfit-m6}
\end{figure}

\begin{figure}[!htb]
    \centering
    \begin{subfigure}
        \centering
        \includegraphics[width=0.65\linewidth]{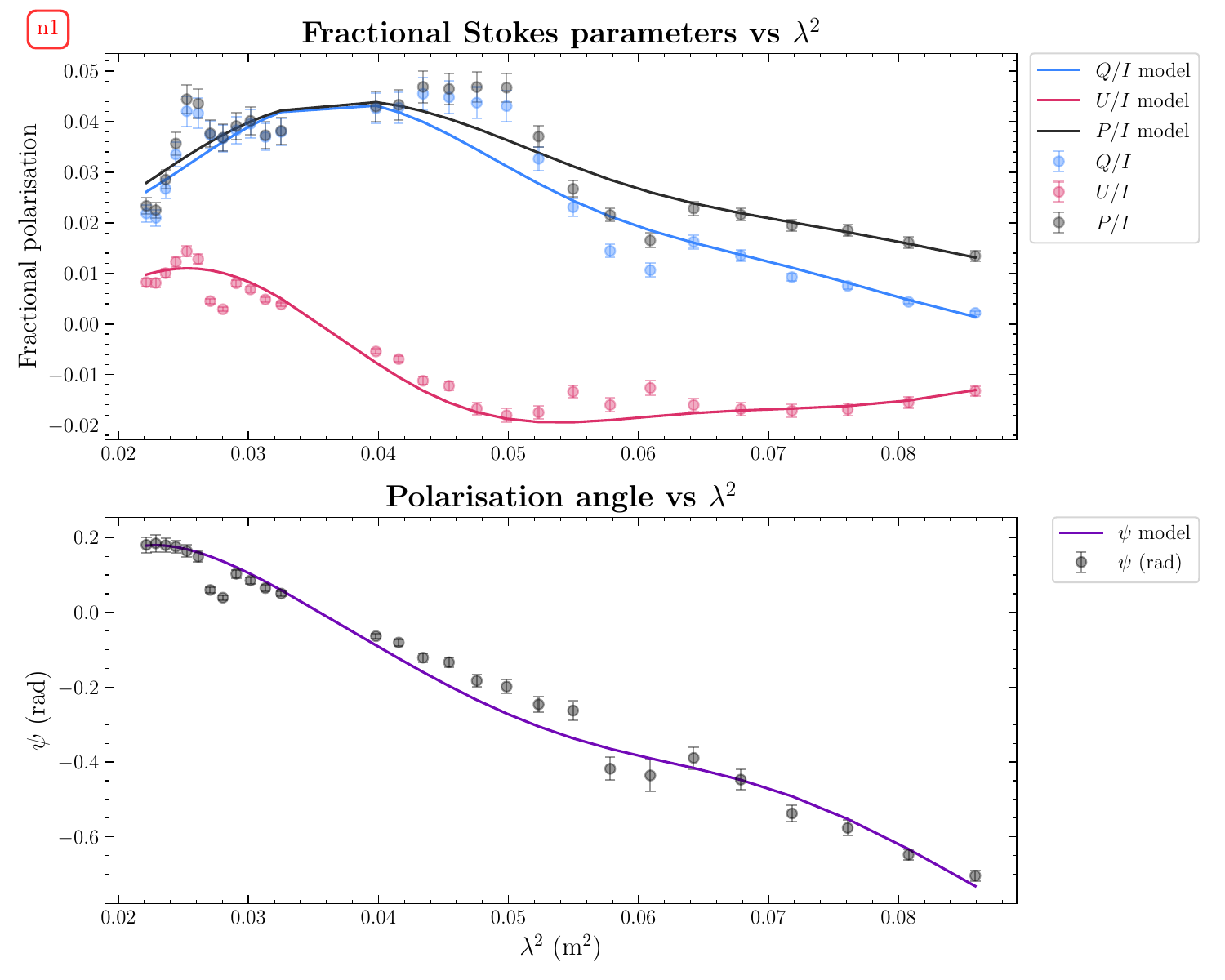}
    \end{subfigure}
    \begin{subfigure}
        \centering
        \includegraphics[width=0.725\linewidth]{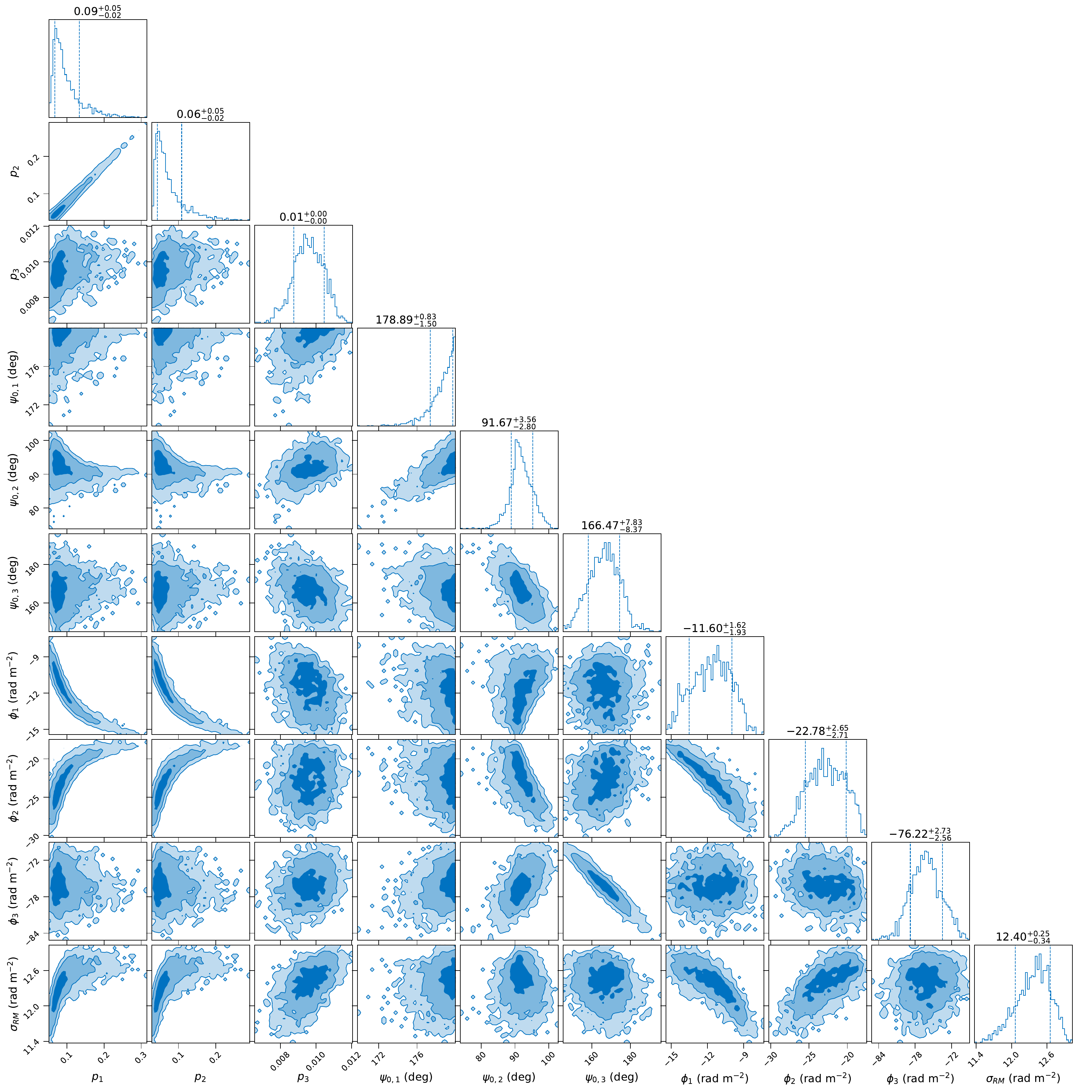}
    \end{subfigure}    
    \caption{QU fitting plots for model n1. Top: Best fit of the fractional polarisation (P/I, Q/I, and U/I) and polarisation angles as a function of $\lambda^2$. The data points represent the observations;   the straight lines are the model prediction (P/I: black, Q/I: blue, U/I: red, $\psi$: purple). Bottom: Probability distributions of all the model parameters: fractional polarisation ($p_1$, $p_2$, and $p_3$), intrinsic polarisation angle ($\psi_{0,1}$, $\psi_{0,2}$, and $\psi_{0,3}$ in deg), Faraday depth ($\phi_1$, $\phi_2$, and $\phi_3$ in $\rm rad \ m^{-2}$), and external Faraday dispersion ($\sigma_{\rm RM}$ in $\rm rad \ m^{-2}$).}
    \label{fig:QUfit-m8}
\end{figure}

\begin{figure}[!htb]
    \centering
    \begin{subfigure}
        \centering
        \includegraphics[width=0.65\linewidth]{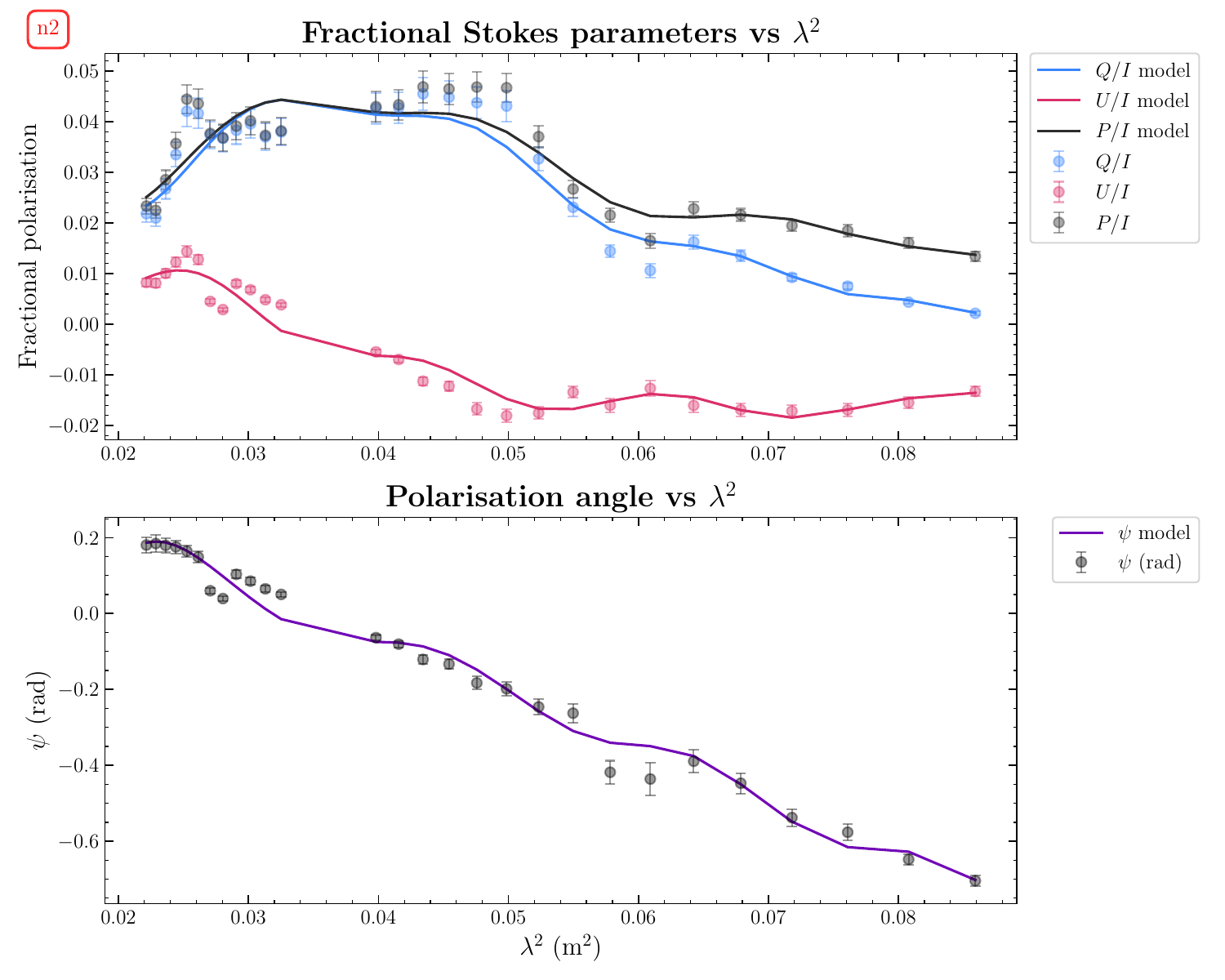}
    \end{subfigure}
    \begin{subfigure}
        \centering
        \includegraphics[width=0.725\linewidth]{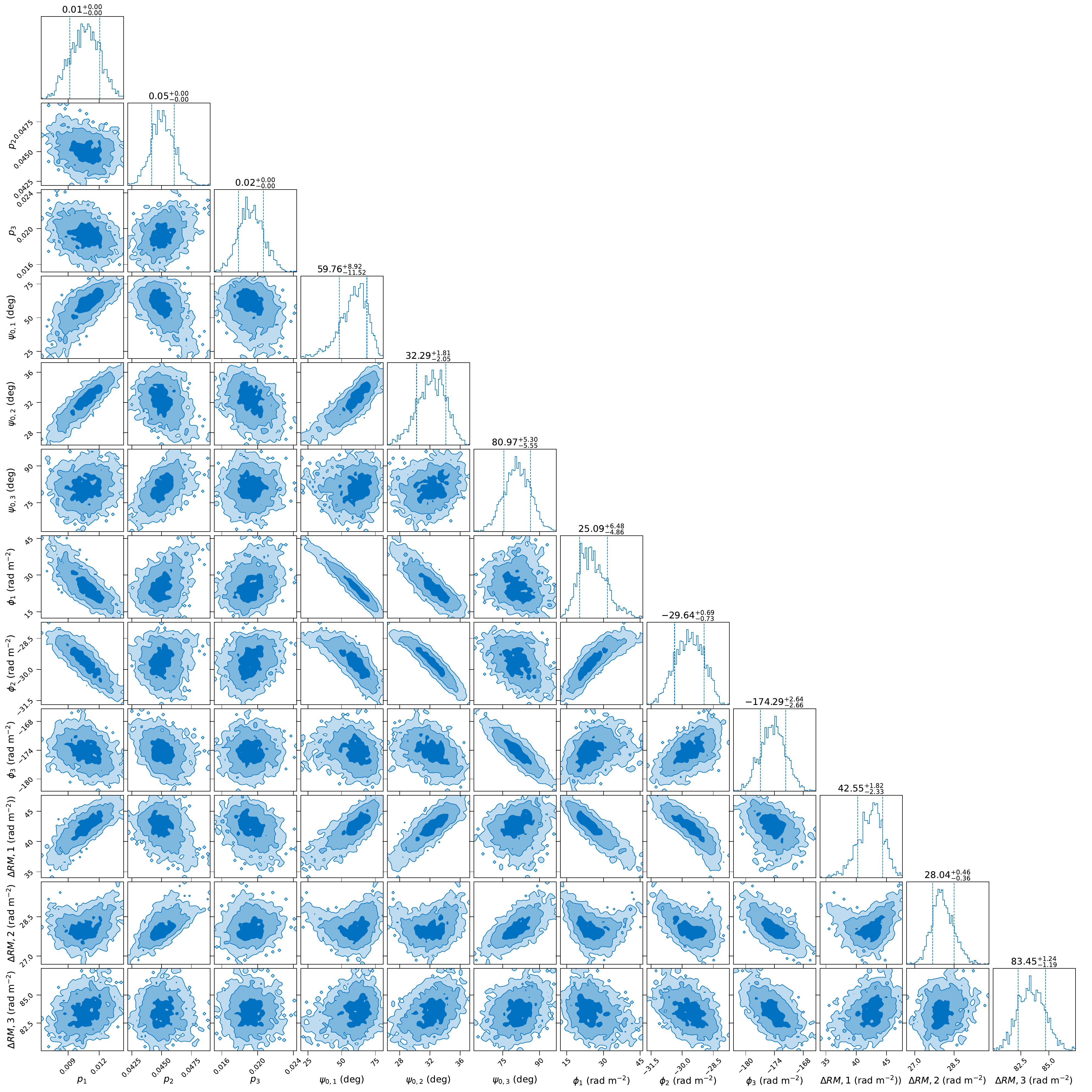}
    \end{subfigure}    
    \caption{QU fitting plots for model n2. Top: Best fit of the fractional polarisation (P/I, Q/I, and U/I) and polarisation angles as a function of $\lambda^2$. The data points represent the observations;  the straight lines are the model prediction (P/I: black, Q/I: blue, U/I: red, $\psi$: purple). Bottom: Probability distributions of all the model parameters: fractional polarisation ($p_1$, $p_2$, and $p_3$), intrinsic polarisation angle ($\psi_{0,1}$, $\psi_{0,2}$, and $\psi_{0,3}$ in deg), mean RM ($\phi_1$, $\phi_2$, and $\phi_3$ in $\rm rad \ m^{-2}$), and internal Faraday dispersion ($\Delta \rm RM_1$, $\Delta \rm RM_2$, and $\Delta \rm RM_3$ in $\rm rad \ m^{-2}$).}
    \label{fig:QUfit-m13}
\end{figure}

\begin{figure}[!htb]
    \centering
        \includegraphics[width=0.725\linewidth]{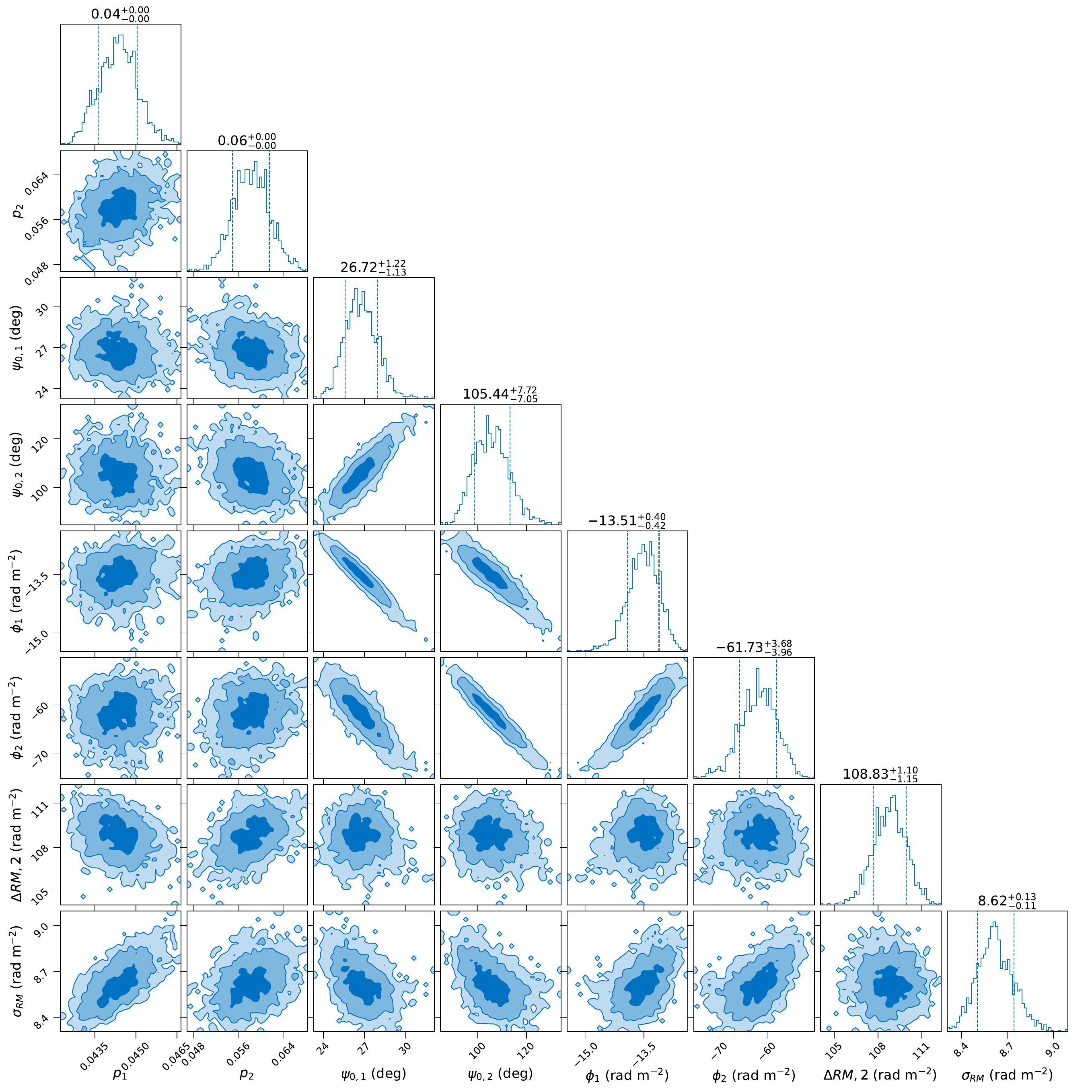}   
    \caption{Probability distributions of all the model parameters for model n3: fractional polarisation ($p_1$ and $p_2$), intrinsic polarisation angle ($\psi_{0,1}$ and $\psi_{0,2}$ in deg), RM of one component ($\phi_1$ in $\rm rad \ m^{-2}$), mean RM of the other component ($\phi_2$ in $\rm rad \ m^{-2}$), internal Faraday dispersion ($\rm \Delta RM_2$ in $\rm rad \ m^{-2}$), and external Faraday dispersion ($\sigma_{\rm RM}$ in $\rm rad \ m^{-2}$).}
    \label{fig:CP-n3}
\end{figure}

In this section, we present the spectral fits and corner plots corresponding to some QU fitting models tested in Sect. \ref{QU fit results}: m2 (Fig. \ref{fig:QUfit-m2}), m3 (Fig. \ref{fig:QUfit-m3}), m6 (Fig. \ref{fig:QUfit-m6}), n1 (Fig. \ref{fig:QUfit-m8}), and n2 (Fig. \ref{fig:QUfit-m13}). We also add the corner plot of model n3 (Fig. \ref{fig:CP-n3}). Spectral plots display the model predictions compared to the observed fractional Stokes parameters Q/I, U/I, and P/I, as well as to the polarisation angle $\psi$, as a function of $\lambda^2$. The corner plots show the posterior distributions of the model parameters derived from the sampling.
\vspace{0.3cm}
\clearpage
\clearpage
\vspace{-0.5cm}
\section{Mock RM maps} \label{MockRM}
\vspace{-0.3cm}
\begin{figure}[!htb]
    \centering
    \hspace{-0.2cm}
    \begin{subfigure}
        \centering
        \includegraphics[width=0.3425\linewidth, height=0.385\linewidth]{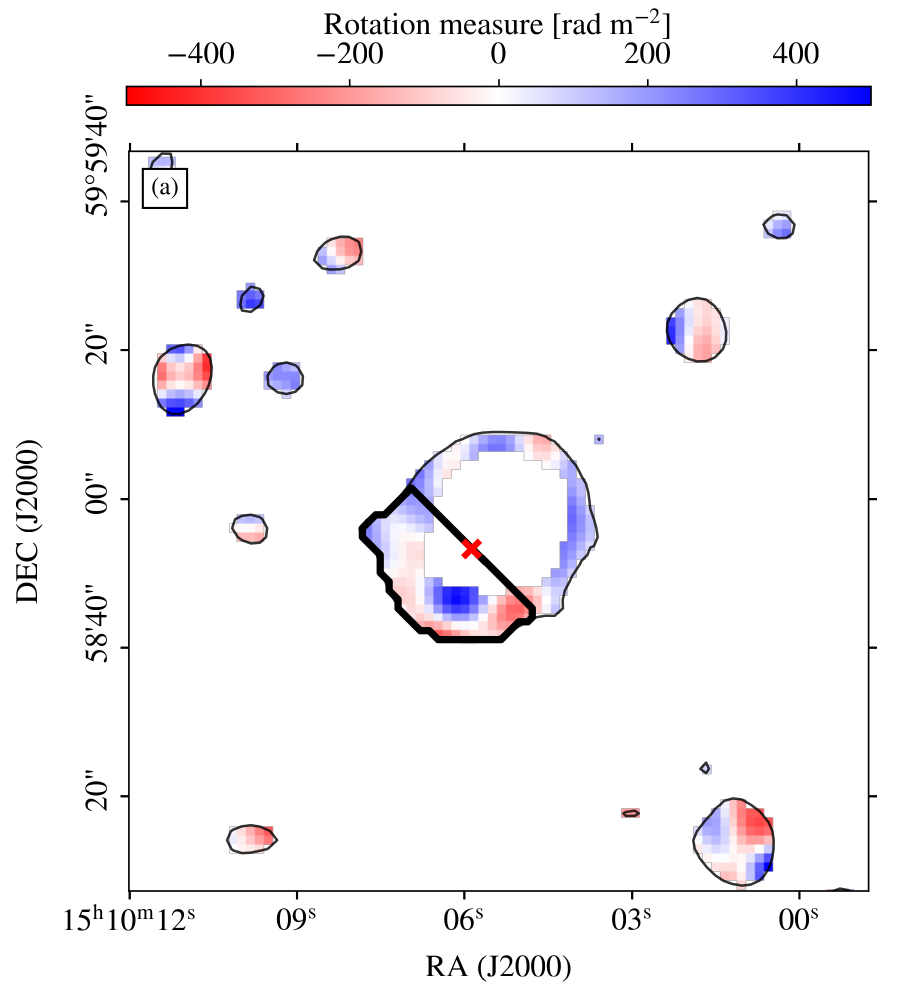}
    \end{subfigure}
    \hspace{-0.4cm}
    \begin{subfigure}
        \centering
        \includegraphics[width=0.34\linewidth, height=0.389\linewidth]{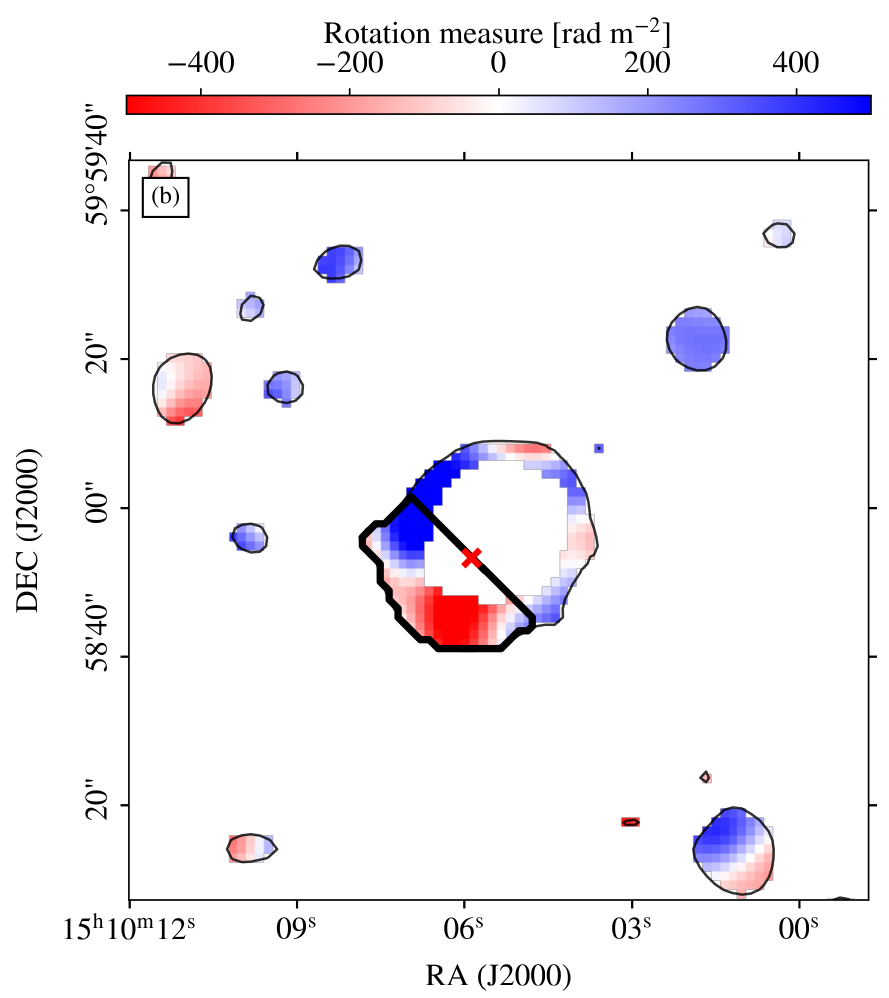}
    \end{subfigure}
    \hspace{-0.4cm}
    \begin{subfigure}
        \centering
        \includegraphics[width=0.34\linewidth,  height=0.3875\linewidth]{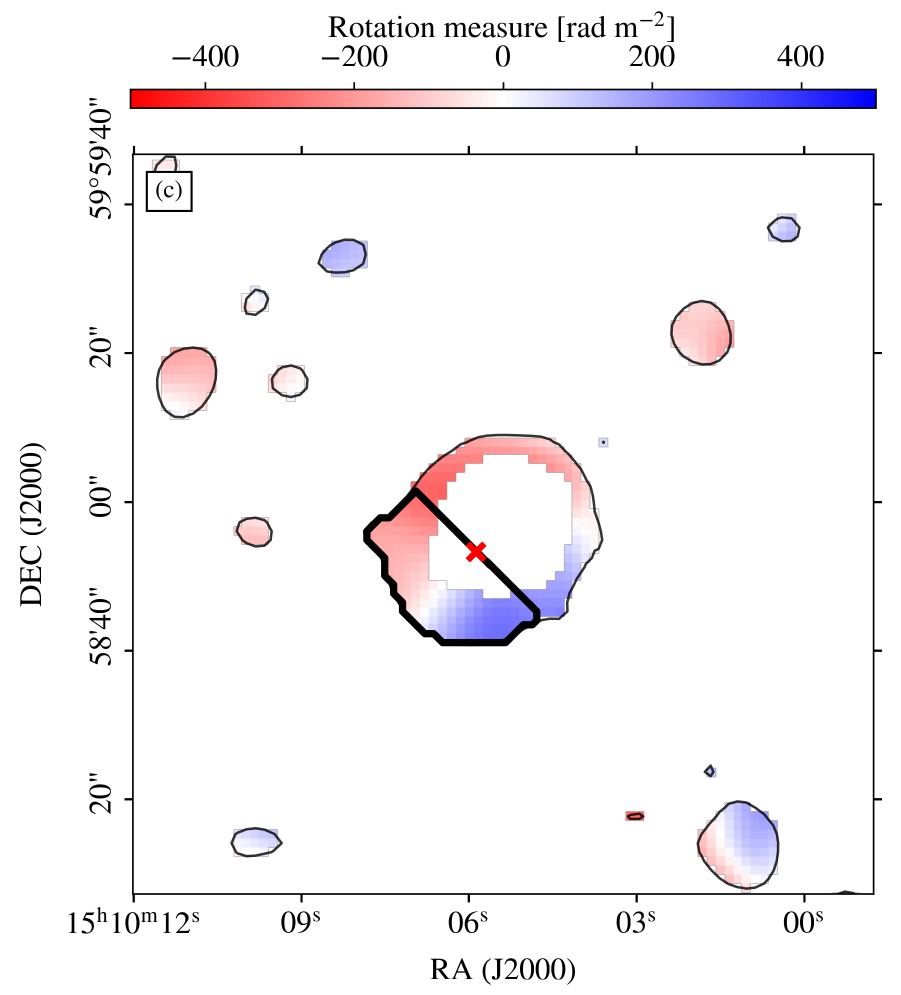}
    \end{subfigure}
    \hspace{-0.4cm}
    \vspace{-0.2cm}
    \caption{Three examples of mock RM maps masked below $\rm 6\sigma_{P,i}$ and above $\rm 3\sigma_{I}$. The thin black contours show the $\rm 3\sigma_I$ level, while the red cross represents the optical position of the host galaxy \citep{Gaia_2021}. The thick black line highlights the region used for the extraction of the RM dispersion. Panel (a): Inhomogeneous case (170 kpc integration length), with $\Lambda_{\rm max} = 102$ kpc, and $\rm B_{\rm norm} = 4 \ \mu{\rm G}$. Panel (b): Homogeneous case ($45^\circ$; 340 kpc integration length), with $\Lambda_{\rm max} = 255$ kpc, and $\rm B_{\rm norm} = 1 \ \mu{\rm G}$. Panel (c): Homogeneous case ($30^\circ$; 600 kpc integration length), with $\Lambda_{\rm max} = 510$ kpc, and $\rm B_{\rm norm} = 0.4\ \mu{\rm G}$.}
    \label{fig:MockRM}
    \vspace{-0.5cm}
\end{figure}
\end{appendix}
\end{document}